\documentclass[a4paper,12pt,twoside]{article}

\usepackage[british]{babel}
\usepackage{url}
\usepackage[colorlinks=true,citecolor=black,linkcolor=black,pagebackref=false,pdftex,urlcolor=blue]{hyperref}

\usepackage{amsmath,amssymb,amsfonts} \usepackage{epsfig, color,
  subfig} \usepackage{enumerate,booktabs,makecell}
\usepackage{changes}
\usepackage{bm}
\usepackage{tikz,tikz-cd}

\addto\captionsbritish{}
\addto\captionsbritish{}
\addto\captionsbritish{}
\addto\captionsbritish{}
\addto\captionsbritish{}
\addto\captionsbritish{}
\addto\captionsbritish{}
\addto\captionsbritish{}
\addto\captionsbritish{}

\definechangesauthor[name=mara, color=blue]{1}
\definechangesauthor[name=marco, color=red]{2}

\def\Ascr{\mathcal{A}}
\def\Bscr{\mathcal{B}}

\def\Gscr{\mathcal{G}}
\def\Hscr{\mathcal{H}}

\def\Lscr{\mathcal{L}}

\def\Oscr{\mathcal{O}}
\def\Pscr{\mathcal{P}}
\def\Sscr{\mathcal{S}}

\def\Pgot{{\mathfrak P}}

\newcommand{\norm}[1]{\left\|#1\right\|}

\def\naturali{\mathbb{N}}
\def\interi{\mathbb{Z}}
\def\reali{{\mathbb{R}}}

\def\toro{\mathbb{T}}

\def\epsilon{\varepsilon}
\def\imunit{{\rm i}}

\def\lie#1{\Lscr_{#1}}

\def\build#1_#2^#3{\mathrel{
\mathop{\kern 0pt#1}\limits_{#2}^{#3}}}

\newtheorem{theorem}{Theorem}[section]
\newtheorem{theorem*}{Theorem}
\newtheorem{lemma}{Lemma}[section]

\title{\bf Existence proof of librational invariant\\ tori in an averaged model\\ of HD60532 planetary system
}

\author{
{\bf VERONICA DANESI}\\
{\small Dipartimento di Matematica,
Universit\`a degli Studi di Roma ``Tor Vergata'',}\\
{\small via della Ricerca Scientifica 1, 00133\ ---\ Roma, Italy.}\\
{\bf UGO LOCATELLI}\\
{\small Dipartimento di Matematica,
Universit\`a degli Studi di Roma ``Tor Vergata'',}\\
{\small via della Ricerca Scientifica 1, 00133\ ---\ Roma, Italy.}\\
{\bf MARCO SANSOTTERA}\\
{\small Dipartimento di Matematica, Universit\`a degli Studi di Milano,}\\
{\small via Saldini 50, 20133\ ---\ Milano, Italy.}\\
{\small e-mails:
  {\tt danesi@mat.uniroma2.it, locatell@mat.uniroma2.it,}}\\
{\small {\tt marco.sansottera@unimi.it}}
}

\begin{document}
\maketitle

\markboth{Existence proof of librational invariant tori in an averaged model of HD60532}{V. Danesi, U. Locatelli, M. Sansottera}

\abstract{ We investigate the
    long-term dynamics of HD60532, an extrasolar system hosting two giant planets orbiting in a 3:1 mean motion resonance. We consider an average approximation  at order one in the masses which results
  (after the reduction of the constants of motion) in a resonant
  Hamiltonian with two libration angles. In this framework, the usual
  algorithms constructing the Kolmogorov normal form approach do not
  easily apply and we need to perform some untrivial preliminary
  operations, in order to adapt the method to this kind of
  problems. First, we perform an average over the
  fast angle of libration which provides an integrable approximation
  of the Hamiltonian. Then, we introduce action-angle variables
  that are adapted to such an integrable approximation. This sequence
  of preliminary operations brings the Hamiltonian in a suitable form
  to successfully start the Kolmogorov normalization scheme. The
  convergence of the KAM algorithm is proved by applying a technique
  based on a computer-assisted proof. This allows us to reconstruct
  the quasi-periodic motion of the system, with initial conditions
  that are compatible with the observations.}

\maketitle


\section{Introduction}
\label{sec:introduction}
The discovery of the first multiple-planet extrasolar
  system, $\upsilon$~Andromed{\ae} (see~\cite{Butler-et-al-1999}),
  immediately raised the question of its stability, notably from a
  dynamical system point of view.  Nowadays more than 800
  multiple-planet extrasolar systems have been discovered, making the
  question even more relevant.

  Typically, these systems have been numerically investigated as a
  sort of {\it inverse problem}, prescribing their stability in order
  to determine ranges of possible values of a few orbital elements
  which are unknown or poorly known (e.g., inclinations and longitudes
  of the nodes).  The numerical investigations of the dynamical
  behavior of many interesting extrasolar planetary systems have been
  done complementing long-term integrations (see,
  e.g.,~\cite{McArt-et-al-2010} and~\cite{Deitrick-et-al-2015}) with
  refined numerical techniques, like for instance the frequency
  analysis method or the MEGNO chaos indicator (see,
  e.g.,~\cite{Las-Cor-2009} and~\cite{Vol-Roi-Lib-2019},
  respectively).

Perturbation theory allows to complement the numerical investigations
with rigorous analytic results.  Normal form methods have a
long-standing tradition and their applications to problems that are
relevant in Celestial Mechanics have grown more and more with the
development of the algebraic manipulators (for an introduction to the
main concepts of this kind of software see,
e.g.,~\cite{Gio-San-Chronos-2012}). Therefore, in such a framework the
study of extrasolar planetary systems (in particular, of their secular
dynamics) started very soon (see, e.g.,~\cite{Mich-Mal-2004}).  The
analytic investigation via computer algebra complements the knowledge
provided by long-term numerical integrations.  In particular, we think
that the modern Hamiltonian perturbation theory gives a proper
framework, where it is possible to naturally explain {\it why a
  planetary configuration is stable} and answer this question also
with quantitative arguments. In this respect, such a goal of the
normal form approaches is somehow reminiscent of the aims of other
recent works about the planetary system dynamics, which are not
limited just to detection of chaos, but they succeed in
explaining which is the source of instability in terms of
superposition of a few resonances that are properly determined
(see~\cite{Mog-Las-2022}).

According to the main results for quasi-integrable systems that have
been obtained in the last decades, effective stability\footnote{A
  dynamical system is said to be {\it effectively stable} when the
  time needed to eventually escape from a small region of the phase
  space is proved to largely exceed the expected life-time of such a
  system.} is ensured in the vicinity of an invariant torus by
  applying the KAM theorem jointly with the Birkhoff normal form and,
  eventually, the Nekhoroshev theorem (see~\cite{Mor-Gio-1995} for a
  complete discussion of this strategy, while applications to
  planetary dynamical models are described in~\cite{Gio-Loc-San-2009}
  and~\cite{Gio-Loc-San-2017}).

In turn, the construction of the invariant torus
  through Kolmogorov normal form is more effective if the starting
  Hamiltonian is close to a suitable normal form designed to locate
  another invariant object. For instance,
  a preliminar (partial) construction of the Birkhoff normal form
  allows one to prove the existence of invariant tori which are in the
  neighborhood of a stable equilibrium point and are well
  approximating the orbits of celestial objects for both the secular dynamics of the
  Sun-Jupiter-Saturn system and the Trojan asteroids
  (see~\cite{Loc-Gio-2000} and~\cite{Gab-Jor-Loc-2005}, respectively).
  In the former case, the equilibrium solution corresponds to orbits
  which are both circular and coplanar in the approximation provided
  by the average over the fast angles (up to order two in the masses)
  of the planetary three-body model; such an approach has been used to
  study the {\it inverse problem} concerning the stability of a few
  extrasolar systems in the framework we have sketched above
  (see~\cite{Vol-Loc-San-2018}). In the latter case, the stationary
  solution is represented by one of the equilateral Lagrangian points,
  that are commonly denoted with $L_4\,$, $L_5\,$; moreover, here it
  has been necessary to preliminarly perform also the construction of
  an intermediate invariant torus well approximating each sought
  torus, by using a variant of the Kolmogorov normalization algorithm
  that avoids small translations on the actions at every step of such
  a computational procedure (which is detailed in
  Section~\ref{sec:KAM}).  In all these works, the rate of convergence
  of the normalization algorithm is as faster as the final invariant
  torus is closer to the equilibrium solution, this distance being
  proportional to the norm of the actions, which are properly defined
  with respect to action-angle canonical coordinates that are
  preliminarly introduced in a suitable way. Therefore, these examples
  highlight that there are regions of the phase space which are
  dynamically stable because they are surrounding KAM tori that, in
  turn, are persistent to perturbations due to their vicinity to
  an elliptic equilibrium point.

A strategy that is similar to the previous one (except for some further
refinement) has made possible to fully develop an application of the
KAM theory to the secular dynamics of a three-body model of the
$\upsilon$~Andromed{\ae} planetary system
(see~\cite{CarLSV-2022}). For that problem, first the normal form for
an elliptic torus has been constructed. Afterwards, an intermediate
invariant torus is constructed by performing the already mentioned
variant of the Kolmogorov algorithm designed so as to skip the small
translations at each normalization step (as it is described in
Section~\ref{sec:KAM}). Finally, the classical Kolmogorov algorithm is
proved to converge to the normal form corresponding to the desired
torus. This result can be explained as follows: the secular dynamics
of the three main bodies of the $\upsilon$~Andromed{\ae} planetary
system is stable because it is strictly winding around a linearly
stable periodic orbit (i.e., a one-dimensional elliptic torus).  The
distance from the elliptic torus to the orbit under consideration
(which is measured with respect to the value of a suitable action
coordinate) has been translated in an easy-to-use numerical
criterion evaluating the robustness of planetary configurations.  Such
a numerical indicator has been successfully applied to the study of
the {\it inverse problem} concerning the stability of the
$\upsilon$~Andromed{\ae} planetary system
(see~\cite{Loc-Car-San-Vol-2021}). This kind of numerical exploration
looks to be very suitable for applications to several (similar)
exoplanetary systems and it is subject of some works in progress.\footnote{M.~Volpi,
  U.~Locatelli, C.Caracciolo, M.~Sansottera. {\it In preparation}.} 

As far as we know, an application of KAM theory to
  realistic models of planetary systems in Mean Motion Resonance
(hereafter often replaced with its acronym MMR) is still lacking;
filling this gap is the main motivation of the present work. Let us
recall that a non-negligible fraction of the multiple-planet
extrasolar systems which have been recently discovered are expected to
be in MMR (see ``The Extrasolar Planet Encyclopedia'', {\tt
  http://exoplanet.eu}). A few of them are hosting exoplanets that
move on rather eccentric orbits; usually they have been detected by
using the Radial Velocity method. We focus our attention on the two
exoplanets orbiting around the HD60532 star. We consider their orbital
dynamics in the framework of the same planar model already considered
in~\cite{Las-Cor-2009} and~\cite{SanLib-2019}, where the existence of
quasi-periodic stable motions is shown by applying the methods of
frequency analysis and a basic normal form approach jointly with
numerical integrations, respectively. In both the papers we have just
mentioned, the model is unambiguously shown to be locked in a $3:1$
MMR, which is double in the sense that there are two independent
combinations of angles (including the mean anomalies) which are in a
libration regime. After having performed an average over a fast
revolution angle and the reduction of the angular momentum, the
problem is described by a two degrees of freedom Hamiltonian. Since
the orbits of those exoplanets are rather far from being circular, we
think that it is not appropriate to limit us to an approach based on
expansions up to the second order in the eccentricities (as it has
been successfully done, with different purposes, in~\cite{BatMor-2013}
and ~\cite{Pucacco-2021}). Therefore, in the present work we study an
Hamiltonian model which is defined by suitable expansions in the
canonical coordinates up to a larger order in the eccentricities
(i.e.,~$6$; see Section~\ref{sec:model} for the proper definitions of
these rather standard expansions). At the end of this paper, in
Section~\ref{sec:KAM} we prove the existence of an invariant KAM torus
carrying quasi-periodic motions which are consistent with the orbits
generated by the numerical integrations starting with initial
conditions compatible with the observations. This result of ours is
fully rigorous in the sense that it is completely demonstrated by
using a computer-assisted proof, based on a normal
  form approach (for an introduction to this method see, e.g., the
  Appendixes of~\cite{Car-Loc-2020}). Let us recall that this is not
  the only viable technique in this context; in particular, a careful
  application of the so called \textit{a posteriori} approach has been
  able to prove the existence of KAM tori for values of the small
  parameter $\epsilon$ extremely close to the breakdown threshold in
  the famous case of the standard map\footnote{So
      remarkable performances are also due to the fact that the
      \textit{a posteriori} method tries to determine just the
      parameterization of the invariant torus, whose existence proof is
      aimed at. Therefore, this approach takes profit of the fact that
      the dimension of the problem is reduced, because the equivalent
      of Taylor expansions with respect to the actions in the phase
      space is not considered (see~\cite{Haro-et-al_book_2016} for a
      description of this computer-assisted technique).} (see
  \cite{Fig-Har-Luq-2017}). In the framework of the computer-assisted
  approach we work with, we emphasize that the preliminary
  approximation of the Kolmogorov normal form is fundamental for the
  eventual success of the application of KAM theory.  Indeed, the
convergence to the final sought KAM torus strongly depends on the
accuracy given by the intermediate normal forms. We emphasize that
none of the strategies we have previously sketched (even if they are
used in junction each other) is sufficient to perform the preliminary
operations in such a way to allow the final constructive algorithm to
be convergent.  Therefore, in
Sections~\ref{sec:case-study}--\ref{sec:action-angle} we need to
carefully describe that part of our approach that is new and so
crucial. We stress that the intermediate Hamiltonian acting as a
keystone for our approach is provided by a further average with
respect to one of the librational angles; this is done so as to
produce an integrable approximation of the final Kolmogorov normal
form, after having performed some further (and suitable) canonical
transformations. Let us also recall that an integrable model for the
dynamics of planetary systems in MMR has been derived in another way
in~\cite{Hadden-2019} and it is used for a different analysis with
respect to ours.

We do believe that the whole computational procedure we describe in
the present paper can apply also to extrasolar planetary systems that
are similar to the one orbiting around HD60532. Nevertheless, the
discussion of the generality of the approach goes beyond our scope
and it is deferred to future investigations.


\section{Resonant Hamiltonian model at order one in the masses}
\label{sec:model}

We consider a planar planetary three-body problem, consisting of a
central star having mass $m_0$ and two coplanar planets having masses
$m_1$ and $m_2$. The problem has $6$ degrees of freedom, which can be
reduced to $4$ due to the conservation of the linear momentum.
Introducing the canonical astrocentric variables
  $(\mathbf{\tilde{r}}_1\,,\,\mathbf{\tilde{r}}_2\,,\,\mathbf{r}_1\,,\,\mathbf{r}_2)$,
  $\mathbf{r}_j$ being the coordinates and $\mathbf{\tilde{r}}_j$ the
  conjugate momenta, the Hamiltonian reads
\begin{equation*}
H(\mathbf{\tilde{r}},\mathbf{r})=T^{(0)}(\mathbf{\tilde{r}})+U^{(0)}(\mathbf{r})+T^{(1)}(\mathbf{\tilde{r}})+U^{(1)}(\mathbf{r}) \ ,
\end{equation*}
where
\begin{align*}
  & T^{(0)}(\mathbf{\tilde{r}})=\frac{1}{2}
  \sum_{j=1}^{2}\norm{\mathbf{\tilde{r}}_j}^2
  \left(\frac{1}{m_0}+\frac{1}{m_j}\right)\ ,
  &U^{(0)}(\mathbf{r})&=-\Gscr\sum_{j=1}^{2}
  \frac{m_0 m_j}{\norm{\mathbf{r}_j}} \ ,
  \\
  & T^{(1)}(\mathbf{\tilde{r}})=\frac{\mathbf{\tilde{r}}_1 \cdot
    \mathbf{\tilde{r}}_2}{m_0} \ ,
  &U^{(1)}(\mathbf{r})&=-\Gscr\frac{m_1 m_2}{\norm{\mathbf{r}_1-\mathbf{r}_2}} 
\end{align*}
and $\Gscr$ is the gravitational constant (see,
e.g.,~\cite{Laskar-1989}).  It is convenient to introduce the
Poincar\'e canonical variables
\begin{align*}
  \Lambda_j &= \frac{m_0 m_j}{m_0+m_j} \sqrt{\Gscr (m_0 + m_j) a_j}\ ,
  \quad &\lambda_j&=M_j + \omega_j\ ,\\
  \xi_j&=\sqrt{2\Lambda_j} \sqrt{1\!-\!\sqrt{1\!-\!e_j^2}} \cos(\omega_j)\ ,
  &\eta_j&= -\sqrt{2\Lambda_j} \sqrt{1\!-\!\sqrt{1\!-\!e_j^2}} \sin(\omega_j)\ ,
  \\
\end{align*}
where $a_j$, $e_j$, $M_j$ and $\omega_j$ are the semi-major axis, the
eccentricity, the mean anomaly and the argument of the pericenter of
the $j$-th planet, respectively.  In addition, we also introduce the
translations $L_j=\Lambda_j-\Lambda_j^*$ where $\Lambda_j^*$ is
defined taking into account the corresponding value $a_j^*$ of the
semi-axis which is compatible with the observations.  Expanding the
Hamiltonian in Taylor-Fourier series around the origin of the
variables $(\bm{L},\bm{\xi},\bm{\eta})$, we get
\begin{equation*}
\vcenter{\openup1\jot\halign{
 \hbox {\hfil $\displaystyle {#}$}
&\hbox {\hfil $\displaystyle {#}$\hfil}
&\hbox {$\displaystyle {#}$\hfil}\cr
 H(\bm{L},\bm{\lambda},\bm{\xi},\bm{\eta})&=K(\bm{L})+\mu P(\bm{L},\bm{\lambda},\bm{\xi},\bm{\eta})\cr
 &=\bm{n}^*\cdot \bm{L}+\sum_{j_1=2}^{\infty}h_{j_1,0}^{(Kep)}(\bm L)+\mu\sum_{j_1=0}^{\infty}\sum_{j_2=0}^{\infty}h_{j_1,j_2}^{(P)}(\bm{L},\bm{\lambda},\bm{\xi},\bm{\eta})\ ,\cr
}}
\end{equation*}
where $n_j^*=\sqrt{\Gscr(m_0+m_j)/(a_j^*)^3}$, for $j=1,2$, and
$\mu=\max\{m_1/m_0,m_2/m_0 \}$. The action-angle variables $(\bm
L,\bm\lambda)$ are referred to as the \emph{fast variables} and the cartesian
variables $(\bm \xi,\bm \eta)$ as the \emph{secular ones}.  In particular,
the functions $h_{j_1,0}^{(Kep)}$ of the Keplerian part $K(\bm{L})$
are homogeneous polynomials of degree $j_1$ in the actions $\bm L$,
while the terms $h_{j_1,j_2}^{(P)}$ of the perturbation
$P(\bm{L},\bm{\lambda},\bm{\xi},\bm{\eta})$ are homogeneous
polynomials of degree $j_1$ in $\bm L$, degree $j_2$ in
the secular variables $(\bm \xi,\bm \eta)$ and trigonometric
polynomials in the angles $\bm \lambda$.

Of course, in practical applications a finite truncation of the
Hamiltonian above is in order. The truncation rules adopted in the present work will be detailed in the
following.


\section{The case study of the HD60532 extra-solar system}\label{sec:case-study}

\begin{table}
\begin{center}
\begin{tabular}{ccccccc}
 \hline
 \rule[-2mm]{0mm}{7mm}
 Planet name & Planet index $j$ & $m_j$ \,[$M_{\text{Jup}}$] & $a_j$\, [AU] & $e_j$ & $\omega_j$ \,[deg] & $M_j\,$ [deg] \\
 \hline
  \rule{0mm}{5mm}
 HD60532b & 1 &  3.1548 & 0.7606 & 0.278 & 352.83 & 21.950 \\
  \rule{0mm}{5mm}
 HD60532c & 2 & 7.4634 & 1.5854 & 0.038 & 119.49 & 197.53 \\
 \hline
\end{tabular}
\end{center}
\caption{Orbital parameters for the HD60532 extra-solar system with
  $i=20^\circ$ and $m_0=1.44$ $M_\odot$, where $M_\odot$ and
  $M_{\text{Jup}}$ stand for the masses of the Sun and Jupiter,
  respectively.}\label{parameters-tab}
\end{table}

\begin{figure}[h]
  \centering
  \includegraphics[scale=0.95]{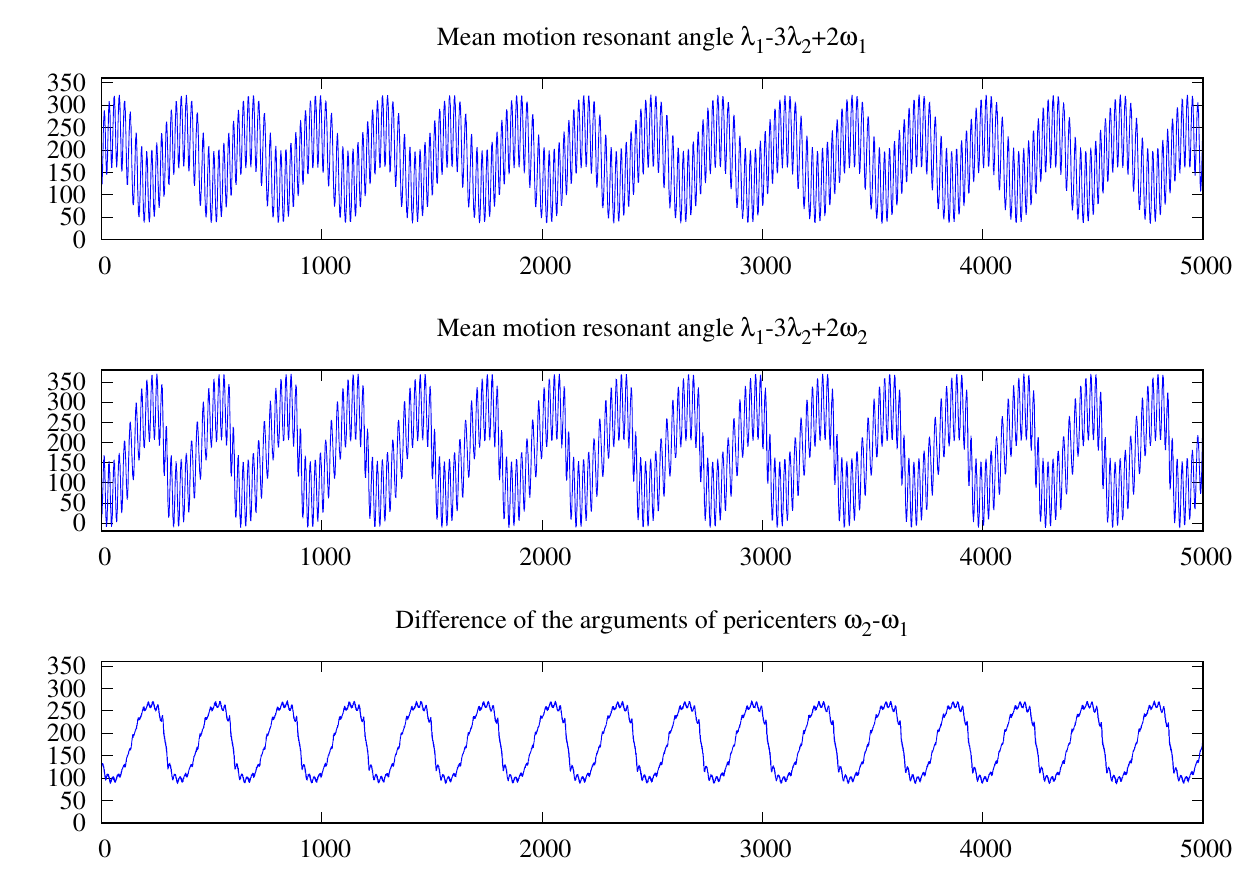}
  \caption{Evolutions in time (whose adopted unit of measure is year)
    of the libration angles, which are measured in degrees and their
    corresponding plots of $\lambda_1-3\lambda_2+2\omega_1\,$,
    $\lambda_1-3\lambda_2+2\omega_2$ and $\omega_2-\omega_1$
    appear in the panels above from top to bottom, respectively.}
  \label{evolutions-res-ang}
\end{figure}

Let us focus on the planar three-body problem for the HD60532
extra-solar system.  The orbital parameters and the initial conditions
are fixed as in Table~\ref{parameters-tab}, according to the values
given in~\cite{Las-Cor-2009,Alv-Mich-Mal-2016,SanLib-2019}.  This
system consists of two giant planets in a $3:1$ MMR, orbiting around
the star named HD60532.  The motion is assumed to be co-planar with an
inclination $i$ (with respect to the plane that is normal to the line
of sight) which is fixed at $20^\circ$.  As a consequence, the initial
masses of the planets are increased by the factor $1/\sin(i)$ with
respect to the minimum ones detected by means of the radial velocity
method.  The presence of the mean motion resonance is confirmed by the
evolution of the resonant angle $\lambda_1-3\lambda_2+2\omega_1\,$,
which librates around $180^\circ$.  Moreover, the system also exhibits
a second libration angle given by the difference of the arguments of
the pericenters $\omega_2-\omega_1\,$, as it has been remarked
in~\cite{Las-Cor-2009,Alv-Mich-Mal-2016,SanLib-2019}. Therefore, also the
average of $\dot\lambda_1-3\dot\lambda_2+2\dot\omega_2$ is equal to
zero.  The evolutions of the resonant angles are reported in
Fig.~\ref{evolutions-res-ang}; they have been produced by running a
symplectic integrator of type $\Sscr\Bscr\Ascr\Bscr_{3}\,$, which is
described in~\cite{Las-Rob-2001}.  The plots of the resonant angles
highlight that the amplitudes of libration are wide, in particular for the resonant angle $\lambda_1-3\lambda_2+2\omega_1\,$,
which has a width of about $280^\circ$.  This makes the study of the
long-term dynamics much more tricky, making it necessary to develop a
suitable approach in order to reconstruct the quasi-periodic motion
pointed out by the numerical integrations of the system. This is the
reason why it is natural to expect that it is convenient to consider
$\lambda_1-3\lambda_2+2\omega_1$ as resonant angle
instead of $\lambda_1-3\lambda_2+2\omega_2\,$, the libration amplitude
of the latter being larger than $360^\circ$.

\begin{figure}[h]
  \centering
  \includegraphics[scale=0.95]{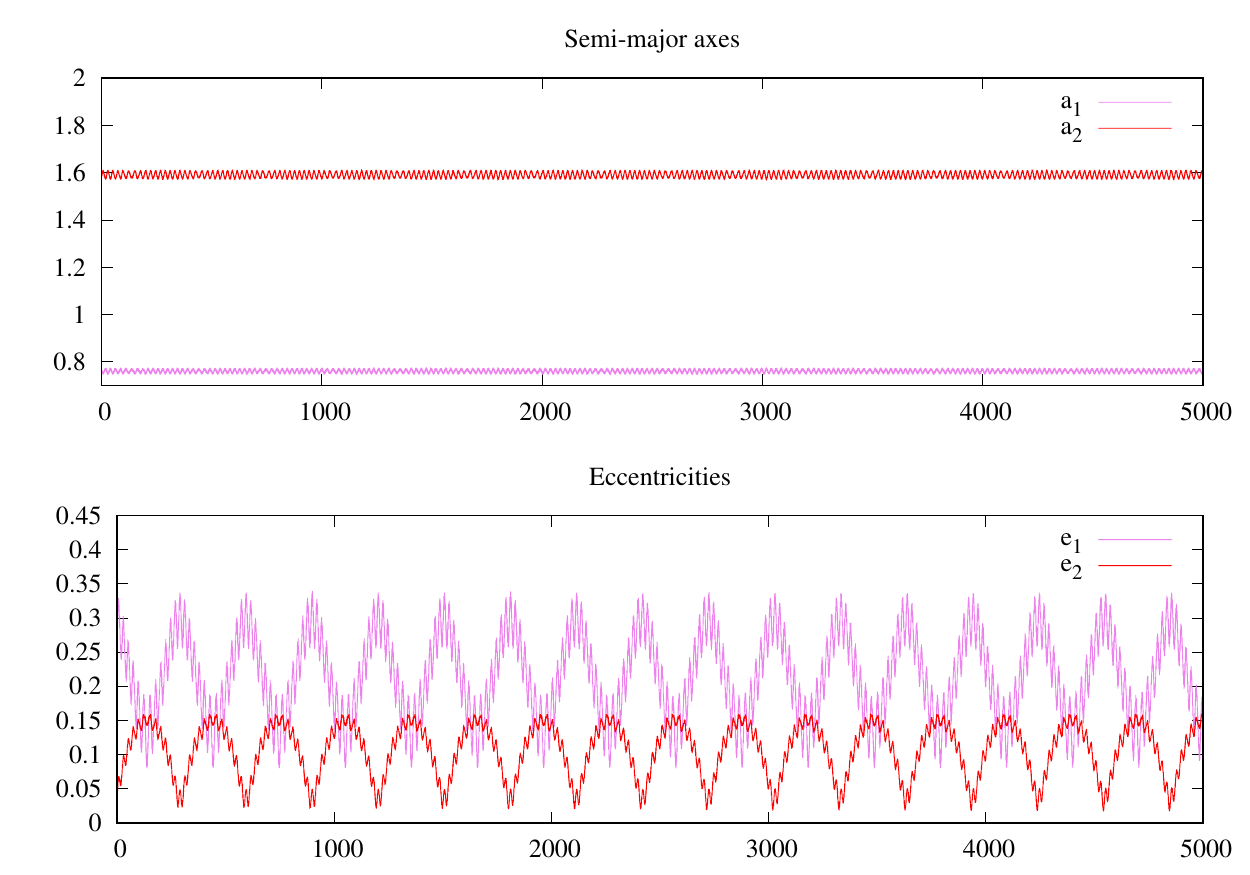}
  \caption{Evolutions in time (measured in years) of the semi-major
    axes [AU] and eccentricities of the exoplanets hosted in the
    HD60532 extra-solar system.}
  \label{evolutions-a+e}
\end{figure}

For what concerns the eccentricities, looking at
Fig.~\ref{evolutions-a+e} one can easily remark that the one of the
inner planet can also exceed the value $0.3$, during its dynamical
evolution. This makes evident that the orbital configuration of these
exoplanets is quite different with respect to that of the biggest
planets of our Solar System, whose orbits are nearly
circular. Therefore, it is natural to expect that a remarkable effort
will be needed to adapt normal forms algorithms which worked
efficiently to construct quasi-periodic approximations of the orbital
motions of the major planets in our Solar System
(see~\cite{Loc-Gio-2005} and~\cite{Loc-Gio-2007}). In order to
efficiently implement a normal form approach to the HD60532
extra-solar system, we will need to design a few modifications to that
basic scheme. This has to be done in such a way to make it more
similar to the approach that successfully worked in the case of the
$\upsilon$~Andromed{\ae} planetary system (see~\cite{CarLSV-2022}),
which also shows the phenomenon of the librations of the difference of
the pericenters arguments (i.e., the so called apsidal locking) as in
the case under study of HD60532.

\subsection{The resonant model}

Being interested in the long-term dynamics of a system
  that is in MMR, we consider a resonant approximation of the
  Hamiltonian that allows to reduce the number of degrees of freedom
  to 2. Hence, we
now consider a set of coordinates which allows to better highlight
this point.  First of all, let us introduce the action-angle variables
$(\bm I,\bm \omega)$ which replace the secular variables $(\bm \xi,\bm
\eta)$ by means of the following canonical transformation:
$$
\xi_j= \sqrt{2I_j}\cos(\omega_j)\ ,
\qquad \eta_j=-\sqrt{2I_j}\sin(\omega_j)\ ,
\qquad \text{for}\quad j=1,2\ .
$$
Now, we also introduce the resonant variables related to the two
libration angles,
\begin{equation*}
  \left\{ 
  \begin{aligned}
    & p_\delta=I_1+2L_1 \\
    & p_\sigma=L_1 \\
    & p_\phi=I_1+I_2+2L_1 \\
    & p_\theta=L_2+3L_1
  \end{aligned}
  \right.
  \qquad \qquad
  \left\{ 
  \begin{aligned}
    & \delta=\omega_2-\omega_1 \\
    & \sigma=\lambda_1-3\lambda_2+2\omega_1\\
    & \phi=-\omega_2 \\
    & \theta=\lambda_2
  \end{aligned} 
  \right.  
  \quad .
\end{equation*}
In this new set of action-angle coordinates, we consider the average
of the Hamiltonian over the (unique) non-resonant revolution angle
$\theta$, i.e.,
$$
\bar{H} = \frac{1}{2\pi}\int_0^{2\pi}
H\big(p_\delta\,,\,p_\sigma\,,\,p_\phi\,,\,p_\theta\,,\,
\delta\,,\,\sigma\,,\,\phi\,,\,\theta\big)\, {\rm d}\theta\ .
$$
Therefore, the angles $\phi$ and $\theta$ turn out to be cyclic
variables for the Hamiltonian $\bar{H}$.  Indeed, the action
$p_\phi$ is exactly the total angular momentum, which is a constant of
motion for the whole three-body planetary system.  Since we perform
an average of the Hamiltonian with respect to a fast angle of orbital
revolution, then it is usual to refer to $\bar{H}$ as a resonant
approximation at order one in the masses. Such an averaged model shows
two first integrals and can be reduced to two degrees of freedom.
The accuracy of the approximation at order one in the masses is discussed in \cite{Lib-San-2013} and \cite{SanLib-2019}, for 
general $2$D three-body models of exoplanetary systems and for particular cases in mean-motion resonance, respectively. This is made by means of comparisons with the results  provided by both the approximation at order two in the masses and the numerical integrations of the non-averaged system.

\medskip

The center of the librations of the resonant angles $\delta$ and
$\sigma$ corresponds to an equilibrium point of the angle variables of
the resonant Hamiltonian $\bar{H}$.  With the aim of expanding
the Hamiltonian $\bar{H}$ around its equilibrium point, we also
look for the values, say $(p_\delta^*,p_\sigma^*)$, of the conjugate
momenta $p_\delta$ and $p_\sigma$ such that the Jacobian of the
Hamiltonian $\bar{H}$ is equal zero.  Once we have determined
the equilibrium point\footnote{For the problem we are considering, we have found the following values: $p_\delta^*=0.0227533$, $p_\sigma^*=-0.00128589$.} $(p_\delta^*,p_\sigma^*,\pi,\pi)$, we can
translate the origin of the canonical variables, by defining
\begin{equation*}
  y_1=p_\delta-p_\delta^*\ , \quad  y_2=p_\sigma-p_\sigma^*\ , \quad
  x_1=\delta-\pi\ , \quad x_2=\sigma-\pi\ ,
\end{equation*}
and expand the Hamiltonian in Taylor series around the origin.  We
also proceed with a diagonalization of the quadratic part of the
Hamiltonian.  Indeed, there is a linear canonical
transformation\footnote{A procedure which allows to determine such a
  canonical transformation $\mathcal{C}$ can be found in Section~7 of
  \cite{GioDFGS-1989}.\label{nota:diagonalization} In order to avoid ambiguities, here the linear transformation $\cal C$ is chosen  in such a way that $|\omega_1| <|\omega_2|$.}
$(y_1,y_2,x_1,x_2)=\mathcal{C}(Y_1,Y_2,X_1,X_2)$ conjugating the
quadratic approximation to a couple of harmonic oscillators.  As a
result, the Hamiltonian in the new polynomial variables $(\bm Y,\bm
X)$ reads
\begin{equation}\label{Ham-diag}
  H(\bm Y,\bm X)=
  \frac{\omega_1}{2}(Y_1^2+X_1^2) + \frac{\omega_2}{2}(Y_2^2+X_2^2) +
  \sum_{\ell \geq 1} h_\ell (\bm Y,\bm X) \ ,
\end{equation} 
where the functions $h_\ell$ are homogeneous polynomials of degree
$\ell+2$ in the variables $(\bm Y,\bm X)$.  Let us remark that,
according to a standard notation in the context of the KAM theory,
hereafter, $\omega_1$ and $\omega_2$ are used to denote the
frequencies (while they have been used before to refer to the
arguments of the pericenters).

The main goal of this work is to investigate the stability of the
Hamiltonian model given by \eqref{Ham-diag} and to reconstruct its
quasi-periodic motion, starting from initial conditions corresponding
to the data reported in Table~\ref{parameters-tab}.  First of all, let
us stress that the Hamiltonian~\eqref{Ham-diag} has an elliptic
equilibrium point at the origin and, in addition, in the case of the
extra-solar system HD60532, the two frequencies $\omega_1$ and
$\omega_2$ also have the same sign.  Hence, it would be quite natural
to try to deal with the problem using a Lyapunov confinement argument
about the values of the actions after having performed a few steps of
the Birkhoff normalization algorithm.  However, this approach fails because the initial conditions (expressed in the polynomial variables $(\bm Y,\bm X)$) are too far from the equilibrium point situated at the origin.
Hence, we need a less naif method in order to tackle the problem under
study.  Therefore, one could try another constructive procedure that
has shown to be successful in a similar context, i.e., for models of
the secular planetary dynamics (see~\cite{Loc-Gio-2000},
\cite{Gio-Loc-San-2017} and~\cite{Vol-Loc-San-2018}). Indeed, it could
be convenient to first introduce action-angle variables, with the aim
of performing a translation of the actions and then applying the
standard Kolmogorov normalization algorithm.  Nevertheless, also this
attempt fails, because it is not enough to achieve the convergence of
the final procedure, even if preceded by a finite number of steps of
the Birkhoff normalization algorithm.

We are then led to develop a different approach which is adapted to
the special kind of problem we are considering.  Let us remark that in
this model a slow dynamics can be distinguished from a faster one, as
we can see from the plots of the two libration angles that are
reported in the first panel of Fig.~\ref{evolutions-res-ang} and the
third one.  In particular, the difference of the argument of the
pericenters points out the slow period, that is $\Oscr(1/\mu)$, while
the mean motion resonant angle $\sigma$ also highlights the presence
of a faster period.  Therefore, the key strategy to face the problem
is to preliminarly average the Hamiltonian with respect to the faster
libration angle, namely over an angle related to the MMR. Let us recall that, by applying the procedure mentioned in
footnote${\ref{nota:diagonalization}}\atop{\phantom{1}}$, it can be
easily shown that the period of such a (so called) {\em fast libration
  angle} is $\Oscr(1/\sqrt{\mu})$. Therefore, it is somehow
intermediate between the secular angles and the orbital revolution
ones. This justifies the name we have decided to adopt, in order to
refer to it.


\section{Average over the fast libration angle}\label{sec:average}

In this section we describe the algorithm which allows to perform the
average of the Hamiltonian with respect to the fast libration angle.

We introduce the action-angle variables $(\bm J,\bm \vartheta)$
via the canonical transformation $(\bm Y,\bm X)=\Ascr(\bm J,\bm
\vartheta)$, namely
\begin{equation}
  \label{action-angle-coord}
  Y_j= \sqrt{2J_j}\cos(\vartheta_j)\ ,
  \qquad X_j=\sqrt{2J_j}\sin(\vartheta_j)\ , \qquad \text{for}\quad j=1,2\ .
\end{equation}
After this canonical change of coordinates the Hamiltonian
\eqref{Ham-diag} reads
\begin{equation}\label{starting_Ham}
  \Hscr^{(0)}(\bm J,\bm \vartheta)=\bm \omega\cdot \bm J
  +\sum_{\ell\geq 1} h^{(0)}_\ell(\bm J,\bm \vartheta)\ ,
  \qquad \text{with} \quad (\bm J,\bm \vartheta)\in\reali^2\times\toro^2 \ ,
\end{equation}
where the functions $h^{(0)}_\ell$ are homogeneous polynomials of
degree $\ell+2$ in the square root of the actions $\bm J$ and
trigonometric polynomials in the angles $\bm \vartheta$.  The
superscript refers to the normalization step of the averaging
algorithm we are going to describe in detail.

\subsection{Formal algorithm for the construction of a resonant Birkhoff normal form}

As usual, this normal form is constructed by using the Lie series
formalism, with the Lie series operator $\exp\left( \lie{\chi}\right)$
defined as follows:
$$
\exp\left( \lie{\chi}\right)=
\displaystyle \sum_{s\geq 0}\dfrac{1}{s!}\lie{\chi}^s
\quad \text{and} \quad \lie\chi\cdot=\left\lbrace\cdot,\chi\right\rbrace \ .
$$
Moreover, we denote by $\Pscr_s$ the class of functions depending
on the action-angle variables $(\bm J,\bm \vartheta)$ in such a way
that, $\forall\ g\in\Pscr_s$, $g\circ\Ascr^{-1}$ is an homogeneous
polynomial of degree $s$ in the cartesian canonical variables $(\bm
Y,\bm X)$. In more detail, the Taylor-Fourier expansion of a
generic function $g\in\Pscr_s$ can be written as
\begin{equation}
 g( \bm J, \bm \vartheta) =\!\!\!
 \sum_{\scriptstyle{{ \bm\ell\in\naturali^{2}}\atop{\ell_1+\ell_2=s}}}\!\!
 \,\sum_{\scriptstyle{{k_1=-\ell_1,\,-\ell_1+2,\ldots,\,\ell_1}\atop{k_2=-\ell_2,\,-\ell_2+2,\ldots,\,\ell_2}}}
 \!\!\!\!\!\!\!\!c_{\bm \ell,\bm k}\,
 \big(\sqrt{J_1}\big)^{\ell_1}\,
 \big(\sqrt{J_2}\big)^{\ell_2}
 \exp\big[\imunit (k_1\vartheta_1+k_2\vartheta_2)\big] \>,
\label{frm:esempio-g-in-PPset}
\end{equation}
where the complex coefficients are such that $c_{\bm \ell,-\bm
  k}={\bar c}_{\bm \ell,\bm k}$. For the sake of brevity, in the
following we will adopt the usual multi-index notation for the powers
in the square roots of the actions, i.e, the product
$(\sqrt{J_1})^{\ell_1}\,(\sqrt{J_2})^{\ell_2}$ will be denoted as
$(\sqrt{\bm J})^{\bm \ell}$; moreover, they will be subject to the
restriction $|\bm \ell|=s$ for every term appearing in the expansion
of a function $g\in\Pscr_s\,$, being $|\bm \ell|:=\ell_1+\ell_2\,$.
In the following Lemma\footnote{Its easy
  proof is sketched (for a wider type of classes of functions) in
  Subsection~3.1 of~\cite{Loc-Car-San-Vol-2022}.}, we are going to
describe the behaviour of such a class of functions with respect to
the Poisson brackets.
\begin{lemma}
  \label{lem:classi-funzioni-Birkh}
  Let $f\in\Pscr_{s_1+2}$ and $g\in\Pscr_{s_2+2}\,$, then $\{f, g\}\in
  \Pscr_{s_1+s_2+2}$ $\forall\ s_1\in\naturali\setminus\{0\}\,$,
  $s_2\in\naturali\setminus\{0\}$.
\end{lemma}

Proceeding in a perturbative way, we want to remove step by step the
dependence on the fast angle $\vartheta_2$ (which is related to the
fast libration angle $\sigma$) from the perturbative part of the
Hamiltonian.  Hence, after having performed $r-1$ canonical changes of
coordinates defined by the Lie series operator, the
Hamiltonian~\eqref{starting_Ham} is brought to the following form:
\begin{equation*}
  \Hscr^{(r-1)}(\bm J,\bm \vartheta)=
  \bm \omega\cdot \bm J+\sum_{\ell= 1}^{r-1}Z_\ell(\bm J,\vartheta_1)
  + \sum_{\ell\geq r} h_\ell^{(r-1)}(\bm J,\bm \vartheta)\ ,
\end{equation*}
where $Z_\ell\in\Pscr_{\ell+2}$ and $h_\ell^{(r-1)}\in\Pscr_{\ell+2}\,$.

Let us remark that, with abuse of notation, we are denoting the new
action-angle variables (that are introduced by the canonical
transformation defined by any normalization step) with the same pair
of symbols $(\bm J,\bm \vartheta)$, which has been used to denote the
arguments of $\Hscr^{(0)}$. As it is usual for the Lie series formalism,
this is done in order to contain the proliferation of the symbols.

The Hamiltonian in normal form up to order $r$ is obtained as
$\Hscr^{(r)}=\exp\left(\lie{\chi_r}\right)\Hscr^{(r-1)}$, where the generating
function $\chi_r$ is determined by solving the homological equation
\begin{equation*}
  \lie{\chi_r} \left(\bm\omega\cdot \bm J\right) +
  h_r^{(r-1)}(\bm J,\bm \vartheta)= Z_r(\bm J,\vartheta_1) \ , 
\end{equation*}
with $Z_r(\bm J,\vartheta_1):=\langle
h^{(r-1)}_r\rangle_{\vartheta_2}$ where as usual
$\langle\cdot\rangle_{\psi}$ denotes the angular average with respect
to $\psi$.  In order to solve such an equation, let us first write the
Taylor-Fourier expansion of the perturbative term as
$$
h_r^{(r-1)}(\bm J,\bm \vartheta)=
 \sum_{\scriptstyle{{ \bm\ell\in\naturali^{2}}\atop{|\bm \ell|=r+2}}}
 \,\sum_{\scriptstyle{{k_1=-\ell_1,\,-\ell_1+2,\ldots,\,\ell_1}\atop{k_2=-\ell_2,\,-\ell_2+2,\ldots,\,\ell_2}}}
 c_{\bm \ell,\bm k}^{(r)}\,
 \big(\sqrt{\bm J}\big)^{\bm \ell}
 \exp\big(\imunit \bm k\cdot \bm \vartheta\big)\ .
$$
Therefore, the $r$-th generating function writes as
\begin{equation*}
  \chi_r(\bm J,\bm \vartheta)=
  \sum_{\scriptstyle{{ \bm\ell\in\naturali^{2}}\atop{|\bm \ell|=r+2}}}
  \,\sum_{\scriptstyle{{k_1=-\ell_1,\,-\ell_1+2,\ldots,\,\ell_1}\atop{k_2=-\ell_2,\,-\ell_2+2,\ldots,\,\ell_2\,;\ k_2\neq 0}}}
  \frac{c_{\bm \ell,\bm k}^{(r)}}{\imunit \bm k \cdot \bm \omega}\,
  \big(\sqrt{\bm J}\big)^{\bm \ell}
  \exp\big(\imunit \bm k\cdot \bm \vartheta\big)\ .
\end{equation*}
Clearly, the generating function can be properly defined if and only
if the frequency vector $\bm\omega$ is non-resonant up to the order
$r+2$.  This means that $\bm k \cdot \bm \omega \neq 0$
$\forall\ 0<|\bm k|\le r+2$. Such a property is certainly satisfied if
we assume that $\bm\omega$ satisfies the Diophantine condition, namely
\begin{equation*}
  |\bm k \cdot \bm \omega|\geq\frac{\gamma}{|\bm k|^\tau}\ ,
  \quad \forall\ \bm k\in\reali^2\setminus \{0\}\ ,
\end{equation*}
for some fixed values of $\gamma>0$ and $\tau\geq 1$. Let us also
recall that almost all the vectors in $\reali^2$ are Diophantine with
respect to the Lebesgue measure.

The transformed functions $h_\ell^{(r)}$ appearing in the expansion of
the new Hamiltonian
\begin{equation}
  \begin{aligned}
    \label{Hamavg-step-r}
  \Hscr^{(r)}(\bm J,\bm \vartheta)&=
  \mathcal{Z}^{(r)}(\bm J,\vartheta_1)+\mathcal{R}^{(r+1)}(\bm J,\bm \vartheta)\\&=
  \bm \omega\cdot \bm J+\sum_{\ell= 1}^{r}Z_\ell(\bm J,\vartheta_1)
  + \sum_{\ell\geq r+1} h_\ell^{(r)}(\bm J,\bm \vartheta)\ ,
  \end{aligned}
\end{equation}
are defined as follows
\begin{equation*}
h_{\ell}^{(r)}= \sum_{j=0}^{\lfloor \ell/r \rfloor} \frac{1}{j!}
                        \lie{\chi_{r}}^{j} h^{(r-1)}_{\ell-jr} \ ,
                  \qquad\qquad  \hbox{for }{\vtop{\hbox{$\ \ell\geq r+1$\ .}
                  }}
\end{equation*}
A simple induction argument, which is based on the application
of Lemma~\ref{lem:classi-funzioni-Birkh}, allows us to verify that
$h^{(r)}_\ell\in \Pscr_{\ell+2}$ $\forall\ \ell$.
Hence, after a finite number $r$ of normalization steps, we get the
Hamiltonian $\Hscr^{(r)}$, which is the sum of a normal form part
$\mathcal{Z}^{(r)}(\bm J,\vartheta_1)=\bm \omega\cdot \bm
J+\sum_{\ell= 1}^{r}Z_\ell(\bm J,\vartheta_1)$, which is integrable,
and a remainder $\mathcal{R}^{(r+1)}(\bm J,\bm
\vartheta)=\sum_{\ell\geq r+1} h_\ell^{(r)}(\bm J,\bm \vartheta)$.
Indeed, the averaged part $\mathcal{Z}^{(r)}(\bm J,\vartheta_1)$ is
independent of the fast angle $\vartheta_2$.  Therefore, the action
$J_2$ is constant along the flow induced by $\mathcal{Z}^{(r)}$,
because $\big\{J_2\,,\,\mathcal{Z}^{(r)}\big\}=0\,$. Moreover, the
averaged part $\mathcal{Z}^{(r)}$ results in an integrable
approximation, because it can be reduced to an Hamiltonian having just
one degree of freedom.

For later convenience, it is also worth to recall that the canonical
transformation $\mathcal{C}^{(r)}$ defining the resonant Birkhoff
normal form up to the $r$-th step of the constructive algorithm is
explicitly given by
\begin{equation}
  \mathcal{C}^{(r)}(\bm J,\bm \vartheta)=
  \exp\lie{\chi_r}\,\circ\,\exp\lie{\chi_{r-1}}
  \,\circ\,\ldots\,\exp\lie{\chi_1}
  \,(\bm J,\bm \vartheta)\ .
\label{eq:trasf_res_Birkh}
\end{equation}
In fact, the exchange theorem for Lie series ensures us that
$\Hscr^{(r)}(\bm J,\bm \vartheta)=\Hscr^{(0)}\big(\mathcal{C}^{(r)}(\bm J,\bm \vartheta)\big)$
$\forall\ (\bm J,\bm \vartheta)\in\Bscr(\bm 0)\times\toro^2$,
being $\Bscr(\bm 0)$ a suitable open ball centered
around the origin of $\reali^2$ (see~\cite{Grobner-60}
and~\cite{Giorgilli-Libro-2022}).

\subsection{Comparison between numerical integrations and semi-analytic solutions}\label{sec:first-comparison}

In this subsection, we are going to check the validity of the averaged
Hamiltonian up to a finite order $\tilde r$ (namely the integrable
approximation $\mathcal{Z}^{(\tilde r)}$) in describing the orbital
motions induced by the Hamiltonian~\eqref{Ham-diag} which describes the slow dynamics of a planetary system in MMR.  For what
concerns our extra-solar model, this latter Hamiltonian, expressed in
action-angle variables as in~\eqref{starting_Ham}, has been expanded
up to order $6$ in the square root of the actions.  We perform $6$
normalization steps and our goal is to compare numerical
integrations\footnote{All the computations discussed in the present
Section and in the following one, which are both of symbolic type and
of purely numerical kind, have been performed by using {\tt
  Mathematica}.} of the Hamiltonian in MMR \eqref{Ham-diag} with the
semi-analytic solution of the averaged Hamiltonian up to order $6$,
both described in the cartesian variables $Y_j=
\sqrt{2J_j}\cos(\vartheta_j)$ and $X_j=\sqrt{2J_j}\sin(\vartheta_j)$,
for $j=1,2$. The choice $\tilde r=6$ allows to obtain
  a reasonable balance between the accuracy and the needed
  computational time.
\begin{figure}
			\centering
        \includegraphics[scale=0.50]{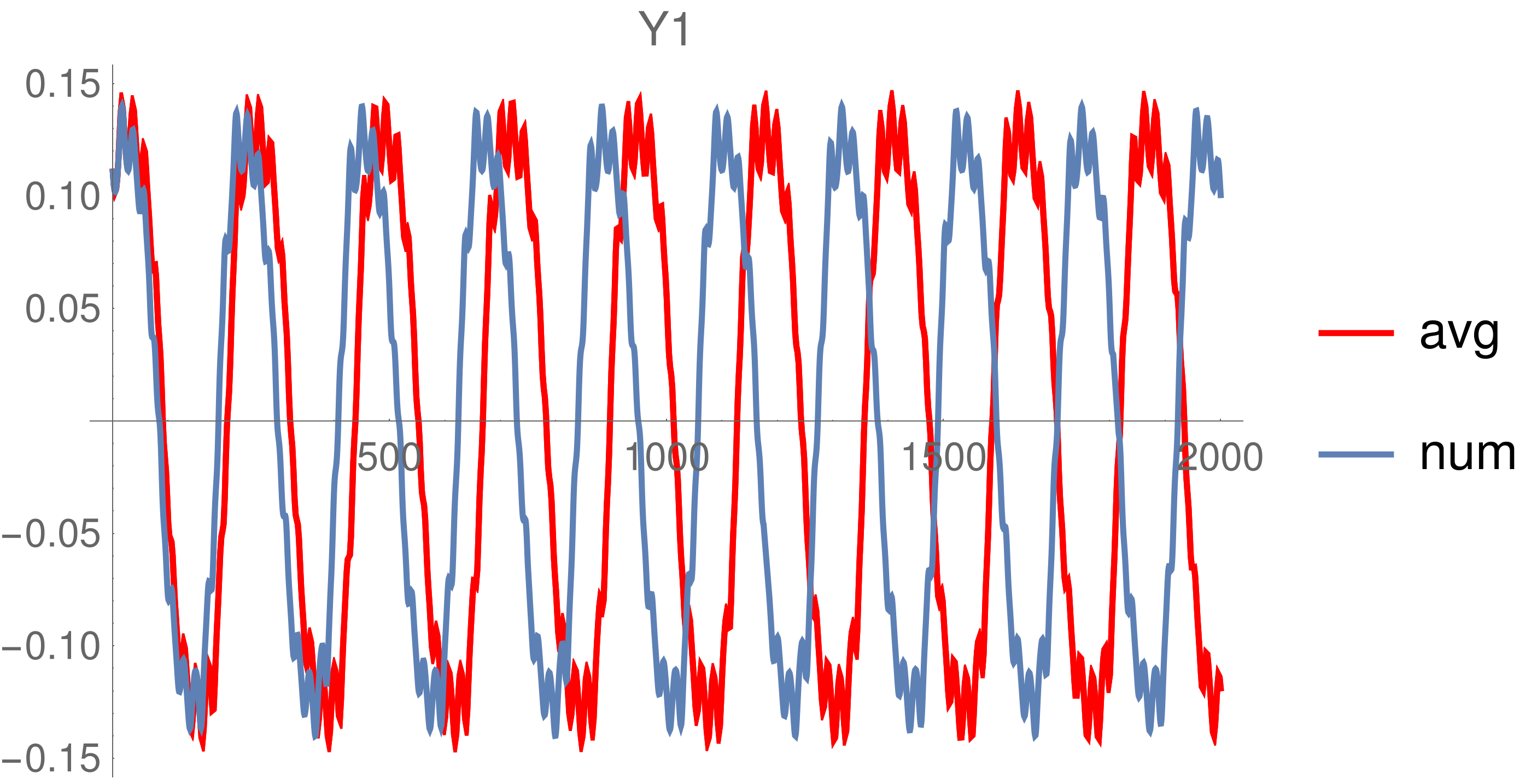}
        \hskip 10pt
        \includegraphics[scale=0.50]{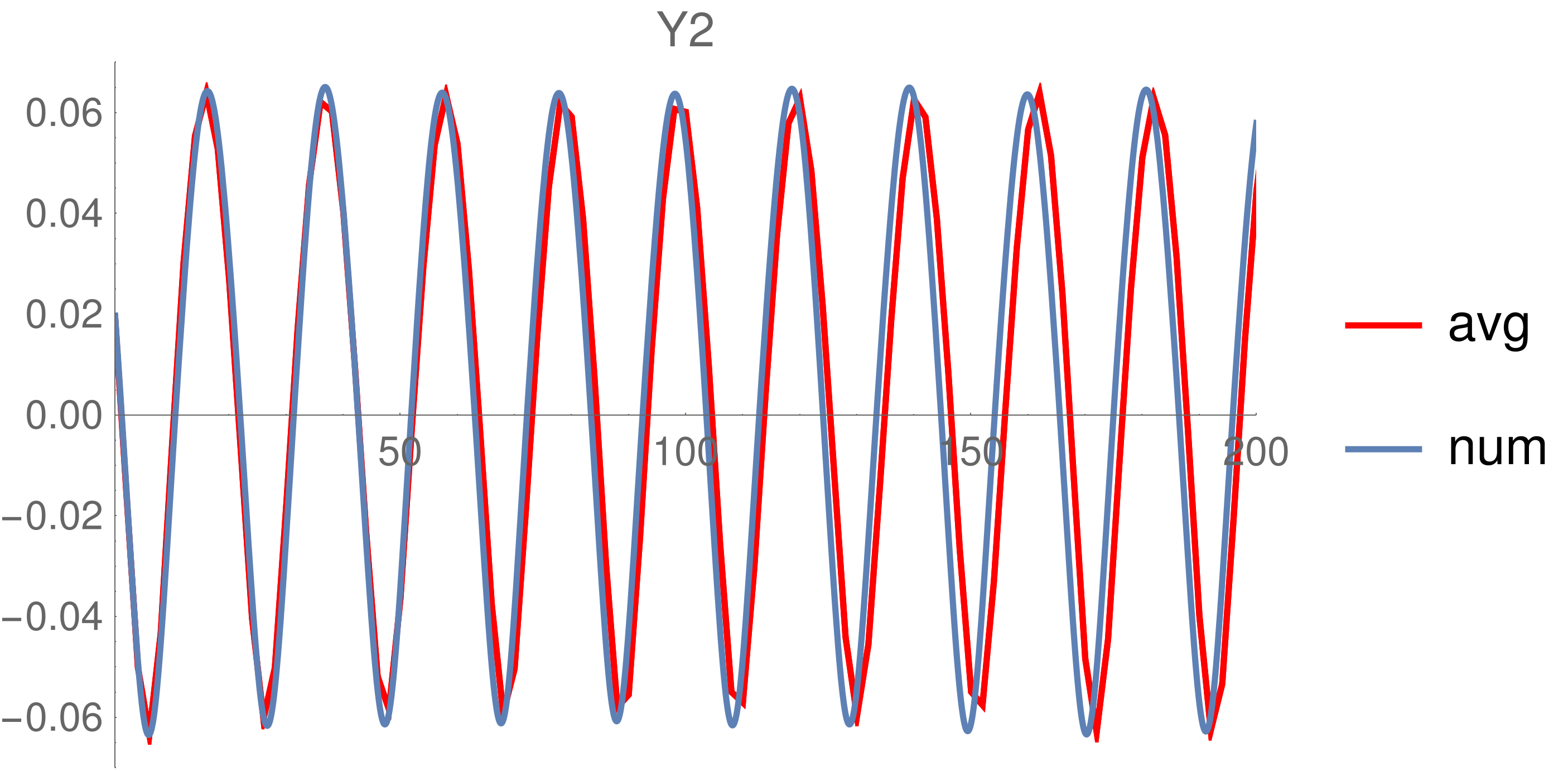}
            \includegraphics[scale=0.50]{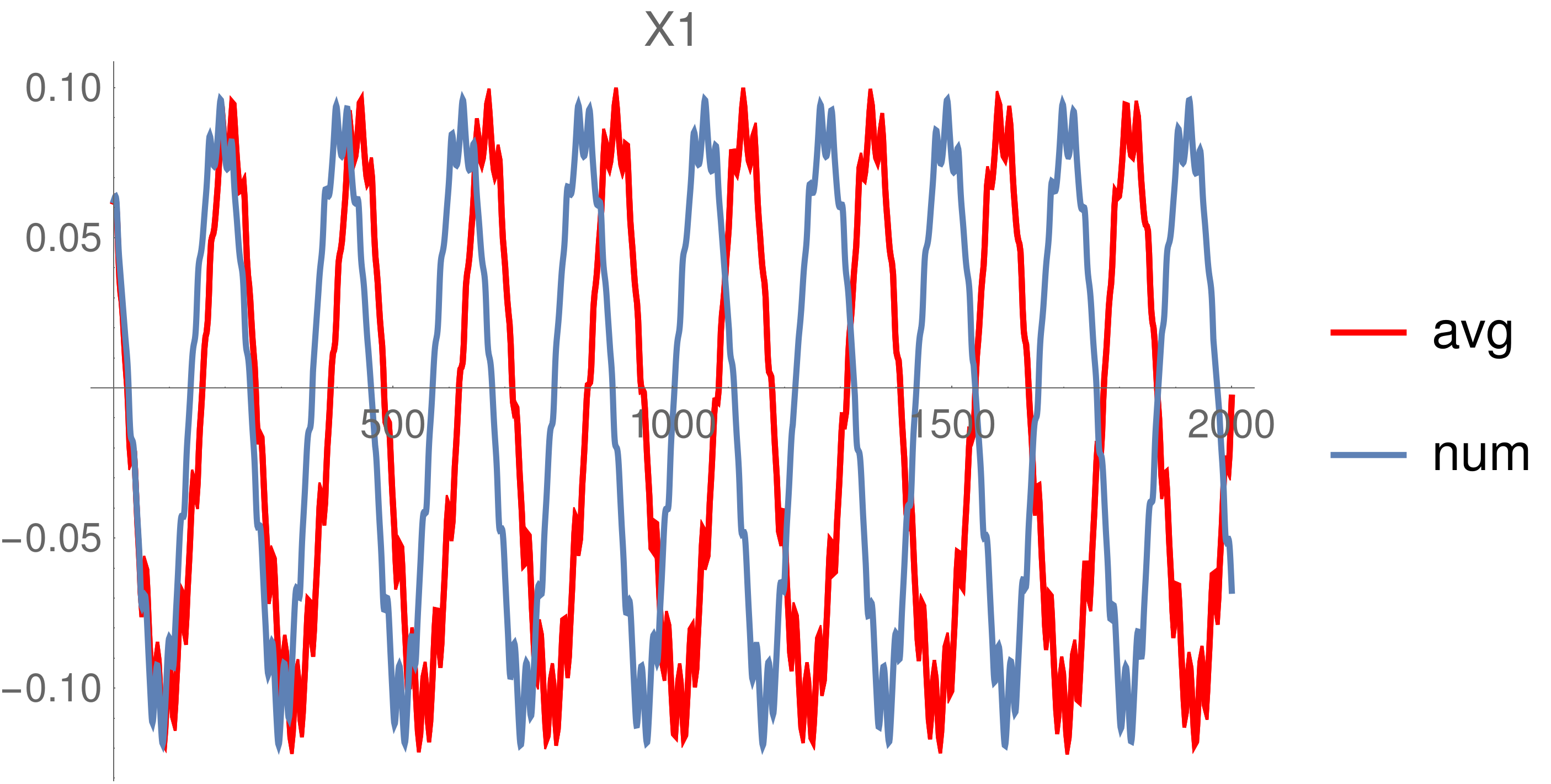}
            \hskip 10pt
        \includegraphics[scale=0.50]{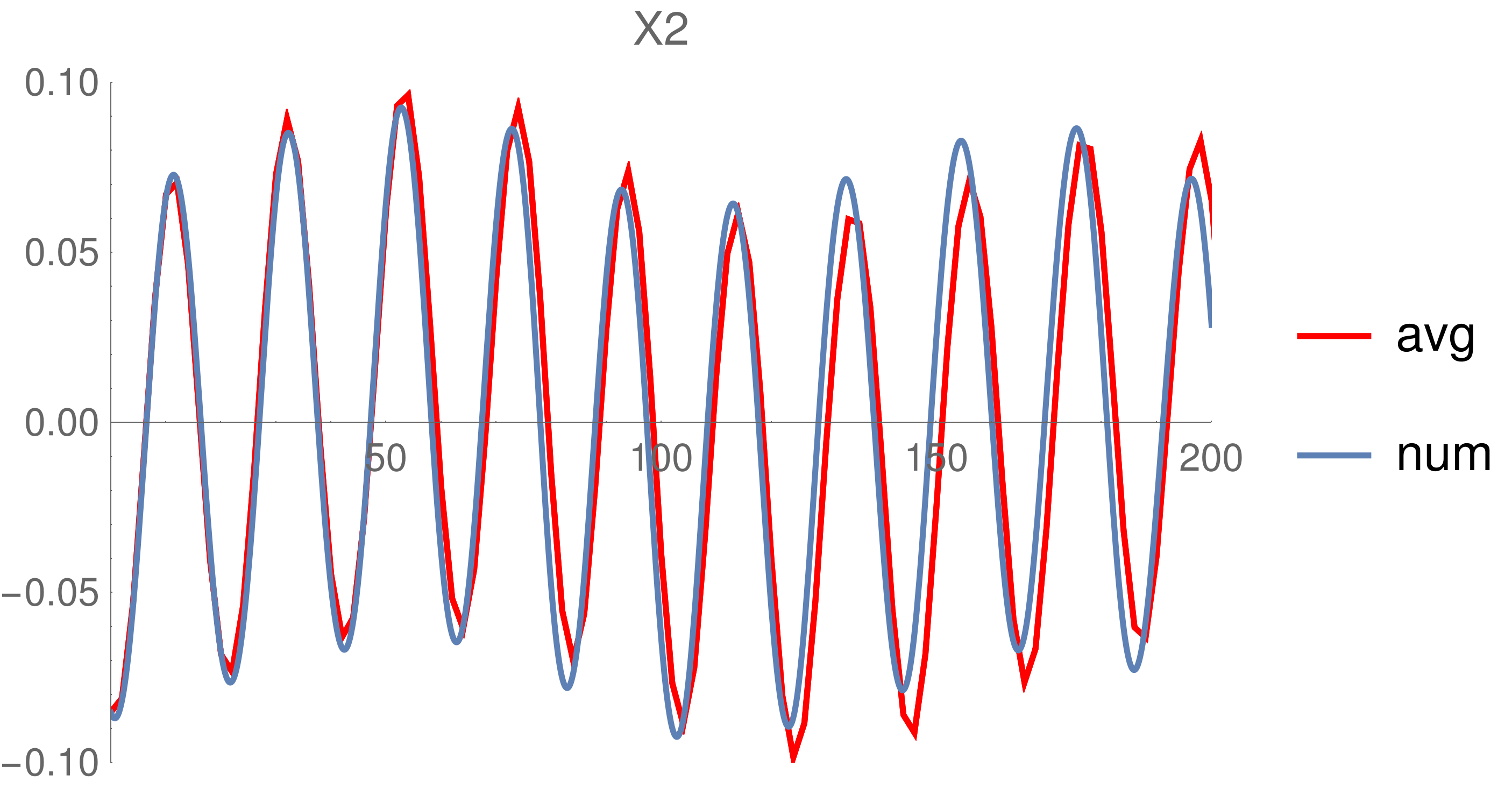}
        \caption{Evolutions in time [yr] of the slow variables (left
          panels) and the fast ones (right panels), given by the
          numerical integrations of the Hamiltonian~\eqref{Ham-diag}
          (blue curves) and the semi-analytic solution of the averaged
          Hamiltonian $\mathcal{Z}^{(6)}$ (red curves).}
        \label{fig:media-num}
\end{figure}

Let us recall that the averaged Hamiltonian $\mathcal{Z}^{(6)}$ is
integrable according to the
Liouville-Arnold-Jost theorem (for a complete proof, see,
e.g.,~\cite{Giorgilli-Libro-2022}).  Therefore, there exists an
analytic expression (eventually involving also the computation of
integrals and the inversion of some functions) which defines a
canonical transformation $(\bm J,\bm \vartheta)=\Psi(\bm P,\bm \varphi)$,
such that the averaged approximation $\mathcal{Z}^{(6)}$ depends on
the actions $\bm P$ only, when it is transformed according to the
change of variables $\Psi$, i.e.,
$$
\frac{\partial\,\mathcal{Z}^{(6)}\big(\Psi(\bm P,\bm \varphi)\big)}
     {\partial\varphi_j}=0
\quad \forall\ j=1,\,2\ .
$$ Thus, in the new set of action-angle variables $(\bm P,\bm
\varphi)$ the equations of motion related to the averaged Hamiltonian
$\mathcal{Z}^{(6)}\circ\Psi$ can be solved very easily.  Moreover, we
can also evaluate the composition $\mathcal{C}^{(6)}$ of canonical
transformations introduced in the previous Subsection in order to
define the action-angle variables $(\bm P,\bm \varphi)$ and to
obtain the Hamiltonian in normal form up to order $6$. The normal form
algorithm can be finally translated in a so called semi-analytic
procedure which allows to determine the motion law $t\mapsto(\bm
Y(t),\bm X(t))$ that is defined by the flow induced by the averaged
Hamiltonian $\mathcal{Z}^{(6)}$. Such a computational procedure is
summarized ($\forall\ t\in\reali$) by the following scheme:
\begin{equation}
\begin{tikzcd}[row sep=5em, column sep=9em,every label/.append style = {font = \normalsize}]
  \big(\bm Y(0),\bm X(0)\big) \arrow[r, "\left(\Ascr\circ\mathcal{C}^{(6)}\circ\Psi\right)^{-1}",shorten <=1em,shorten >=1em] & \big(\bm P(0),\bm \varphi(0)\big) \arrow[d, "\Phi_{\mathcal{Z}^{(6)}\circ\Psi}^t",shorten <=0.5em,shorten >=0.5em] \\
  \big(\bm Y(t),\bm X(t)\big) \arrow[r,leftarrow,"\Ascr\circ\mathcal{C}^{(6)}\circ\Psi",shorten <=0.5em,shorten >=0.5em] & \big(\bm P(t)=\bm P(0),\bm\varphi(t)=\bm\beta t+\bm\varphi(0)\big)
\end{tikzcd}
\label{semi-analytical_scheme}
\end{equation}
where $\Phi_{\mathcal{Z}^{(6)}\circ\Psi}^t$ is nothing but the flow at
time $t$ induced by the Hamiltonian $\mathcal{Z}^{(6)}\circ\Psi$ and
the angular velocity is given by $\beta_j=\frac{\partial}{\partial
  P_j}\big(\mathcal{Z}^{(6)}\circ\Psi\big)$, while $\Ascr$ and
$\mathcal{C}^{(6)}$ are defined in~\eqref{action-angle-coord}
and~\eqref{eq:trasf_res_Birkh}, respectively.  Let us stress that the
initial conditions $(\bm P(0),\bm \varphi(0))$ can be obtained by
inverting the composition of the canonical transformations previously
described, while the initial conditions $(\bm Y(0),\bm X(0))$ are the
ones derived from the observations.  This semi-analytic solution could
be compared with the one obtained by a direct integration of the
Hamiltonian~\eqref{Ham-diag}.  For the sake of simplicity, we do not
perform the last canonical transformation $\Psi$, which is essential
to properly define the semi-analytic
scheme~\eqref{semi-analytical_scheme}, but it would require to perform
some operations (e.g., the aforementioned integrals and the inversions
of functions) that can be hard to implement in a fully explicit way.
We just exploit the uniqueness of the solution of the corresponding
Cauchy problem and we approximate it numerically, by directly
integrating the equations of motion of the averaged Hamiltonian
$\mathcal{Z}^{(6)}$.  Afterwards, we use the canonical
transformations~\eqref{action-angle-coord}
and~\eqref{eq:trasf_res_Birkh} to express the solution in the
variables $(\bm Y(t),\bm X(t))$ and we compare it with the numerical
integration of the Hamiltonian~\eqref{Ham-diag}.

As we can see from the plots in~Fig.\,\ref{fig:media-num}, for the
relatively faster pair of variables $(Y_2,X_2)$ we have a good
agreement between the two solutions, both in terms of amplitude and in
terms of frequency.  Instead, as regards the slow variables
$(Y_1,X_1)$, there is a remarkable error concerning the frequency.
In principle, this discrepancy might be amended with an approximation
  at order two in the masses (which can be adapted to planetary systems in MMR, as explained in~\cite{SanLib-2019}), but this goes beyond the scope of the present paper.


\section{Action-angle variables adapted to the integrable approximation}\label{sec:action-angle}

Before showing that KAM theorem applies in the present context, we
need another preliminary essential step in order to make the algorithm
convergent.  Specifically, we have to introduce a set of action-angle
variables, that are more suitable to describe the integrable
approximation of the Hamiltonian~\eqref{Hamavg-step-r} than the pair
$(\bm J,\bm \vartheta)$ as it is defined after having performed the
canonical transformation $\mathcal{C}^{(6)}$. Indeed, the ideal
action-angle coordinates would be $(\bm P,\bm \varphi)$, the
ones we avoided to compute, because of the technical difficulties due
to an eventual application of the Liouville-Arnold-Jost theorem. Let us
recall that $P_1$ and $P_2$ would be constant of motion for the
integrable approximation $\mathcal{Z}^{(6)}$ and the same holds true
also for the action $J_2\,$.
\begin{figure}[h]
  \centering
  \includegraphics[scale=0.49]{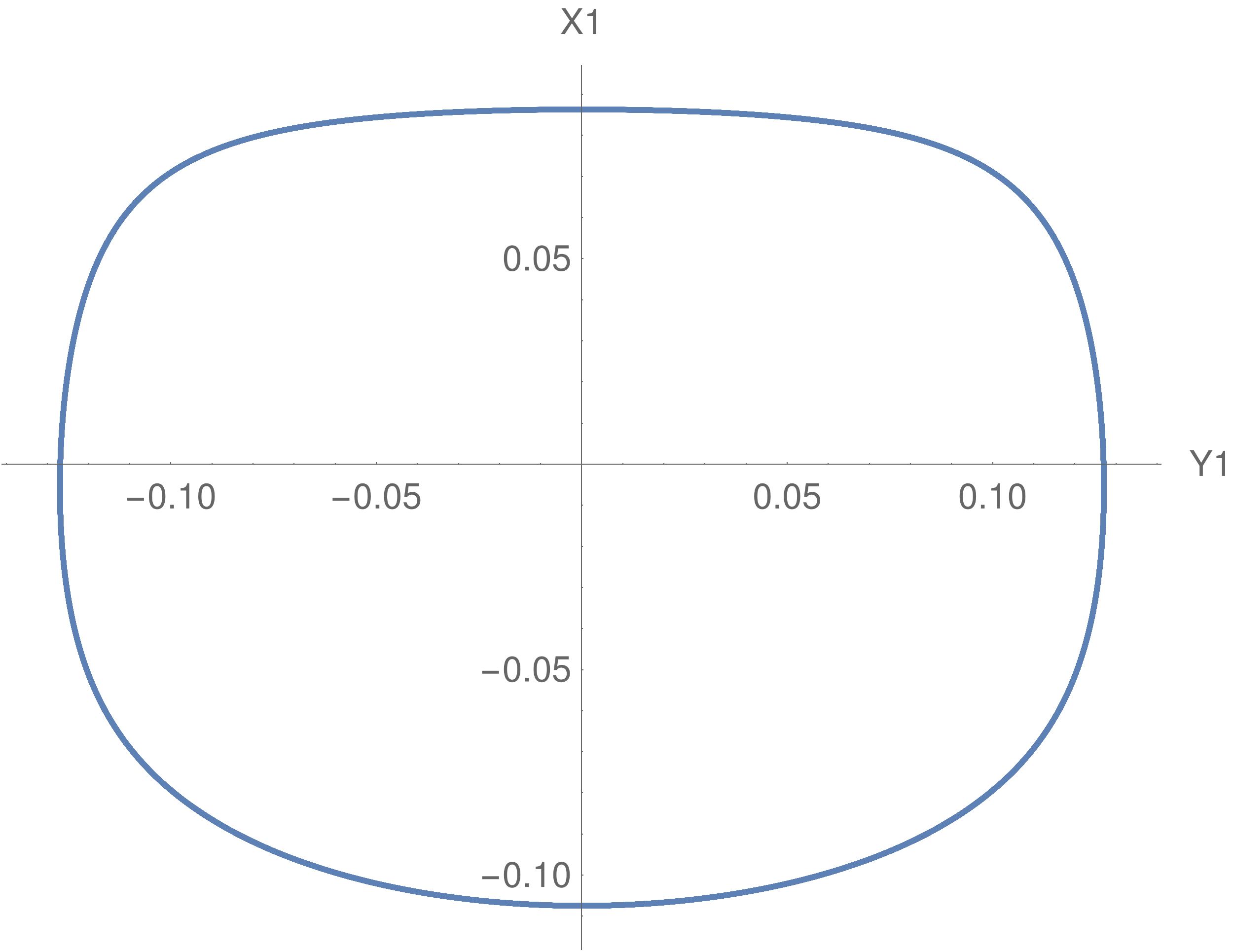}
  \hskip 10 pt
  \includegraphics[scale=0.42]{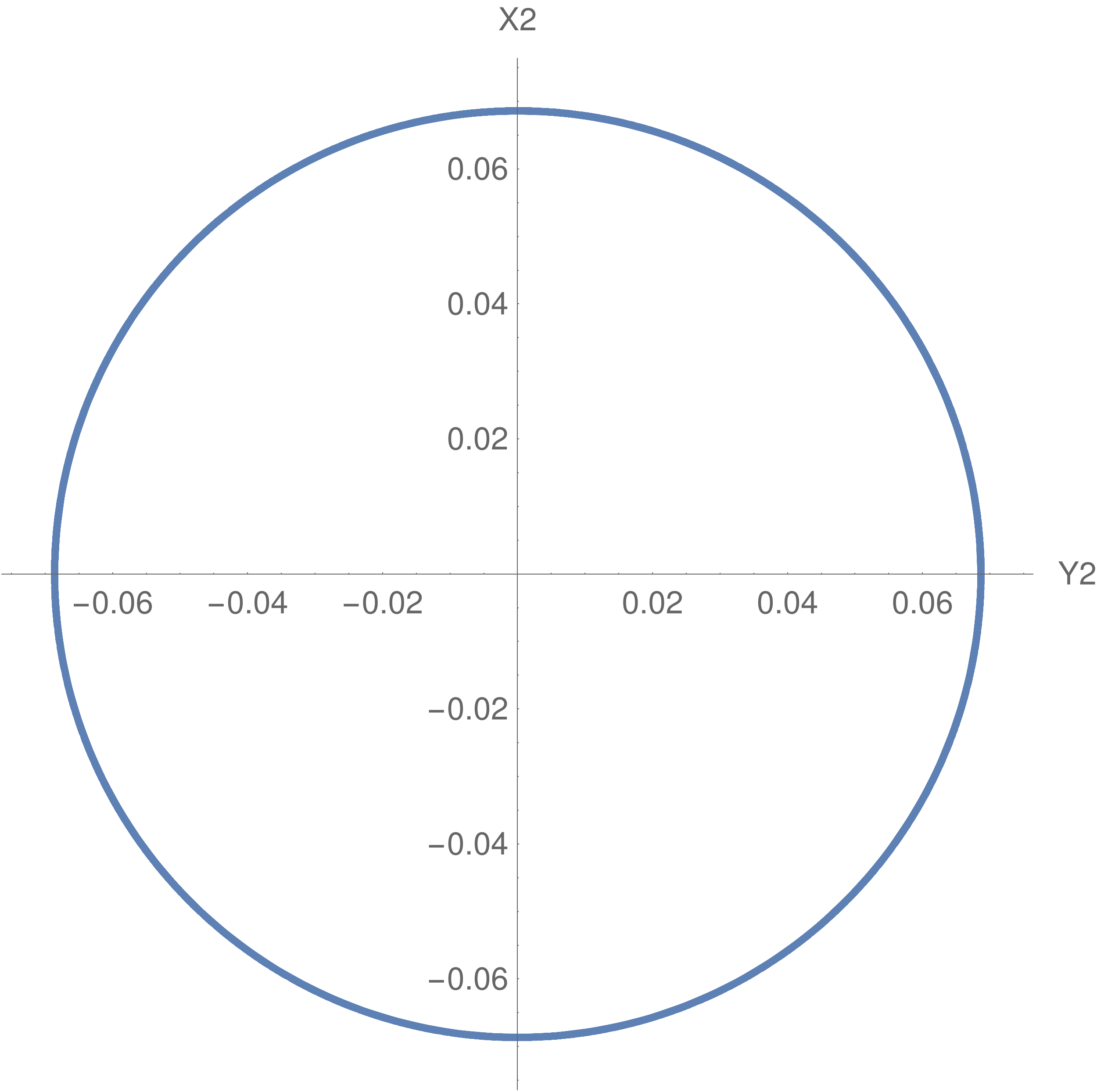}
  \caption{Orbits of the slow dynamics (left panel) and
    the faster one (right panel) for what concerns the
    integrable approximation $\mathcal{Z}^{(6)}$. Both
    the plots refer to the phase space $(\bm Y,\bm X)$.}
  \label{plot_circY1X1Y2X2}
\end{figure}
Hence, if we consider the orbit of the fast motion of the integrable
approximation $\mathcal{Z}^{(6)}$ in the cartesian
variables\footnote{Let us remark that, once again, with a little abuse
  of notation, we are denoting the variables used before and after the
  averaging normalization algorithm with the same name.}
$(Y_2,X_2)=(\sqrt{2J_2}\cos(\vartheta_2),\sqrt{2J_2}\sin(\vartheta_2))$,
we get a circular orbit, as it is shown in the right panel of
Fig.~\ref{plot_circY1X1Y2X2}.  Instead, looking at the orbit of the
slow motion in the cartesian variables
$(Y_1,X_1)=(\sqrt{2J_1}\cos(\vartheta_1),\sqrt{2J_1}\sin(\vartheta_1))$,
that are related to the secular dynamics, we can observe that such an
orbit is far from being circular.  Therefore, at this stage we aim at
introducing a second action which is closer than $J_1$ to be a
constant of motion. In other words, our approach consists in the
construction of action-angle variables with the aim of trying to
circularize (at least partially) the orbit describing the slow
dynamics in the integrable approximation.  Thus, trying to introduce
an action which depends only on the distance from the origin in the
cartesian plane endowed with coordinates $(Y_1,X_1)$, we are dealing
with a quasi-constant of motion and we approach better the sought
(final) KAM torus. From a practical point of view, our approach is
translated in an explicit computational procedure, by suitably
applying the frequency analysis method to the flow induced by the
integrable approximation $\mathcal{Z}^{(6)}$ (see~\cite{Laskar-2003}
for an introduction to such a numerical technique). This can be done
by studying the Fourier decomposition of the signal $Y_1(t)+\imunit
X_1(t)\simeq\sum_{j=1}^{\mathcal{N}_c}A_{1,j} e^{\imunit(k_j\nu_{1}
  t+\varphi_{1,j})} $, where $\mathcal{N}_c$ is the number of
components considered, $A_{1,j}>0$, $k_j\in\interi$,
$\varphi_{1,j}\in(-\pi,\pi]$ and $2\pi/\nu_1$ is the period of such a
motion law.  By taking into consideration only the $\mathcal{N}_c=3$
components of the signal which correspond to $k_j=0,\pm 1$ for
$j=1,2,3$, it is easy to show that the corresponding approximation of
the orbit which describes the secular dynamics is an ellipse.
Therefore, in order to give a circular shape to such an approximation
of the orbit it is necessary to perform two changes of coordinates: a
shift on the variable $X_1$ of a translation value $X_1^*$ and a
dilatation/contraction with coefficient $\alpha$.  The value $X_1^*$
of the translation is determined by exploiting the constant component
(because for $j=1$ we have $k_1=0$ and $\varphi_{1,1}=-\pi/2$, then
$A_{1,j} e^{\imunit\varphi_{1,1}}$ is purely imaginary and so is
aligned with the $X_1$ axis), while the coefficient $\alpha$ is
defined as follows
$$
\alpha=\sqrt{\frac{c_- -c_+}{c_- +c_+}}\ ,
$$
where $c_-$ and $c_+$ are the absolute values of the complex
coefficients of the components with $k_2=-1$ and $k_3=1$,
respectively. In more detail, we define $c_-=A_{1,2}$ and
$c_+=A_{1,3}$ for $j=2,3$; indeed, the easy computation of the
suitable coefficient $\alpha$ of dilatation/contraction takes
profit of the values of the corresponding angles, which are
such that $\varphi_{1,2}=-\varphi_{1,3}\,$. Therefore, we
introduce the new variables
\begin{equation}
  v_1=\alpha\cdot Y_1\ , \qquad u_1=\frac{X_1-X_1^*}{\alpha} \ .
  \label{eq:trasf2morecircular}
\end{equation}

\begin{figure}[h]
  \centering
  \includegraphics[scale=0.45]{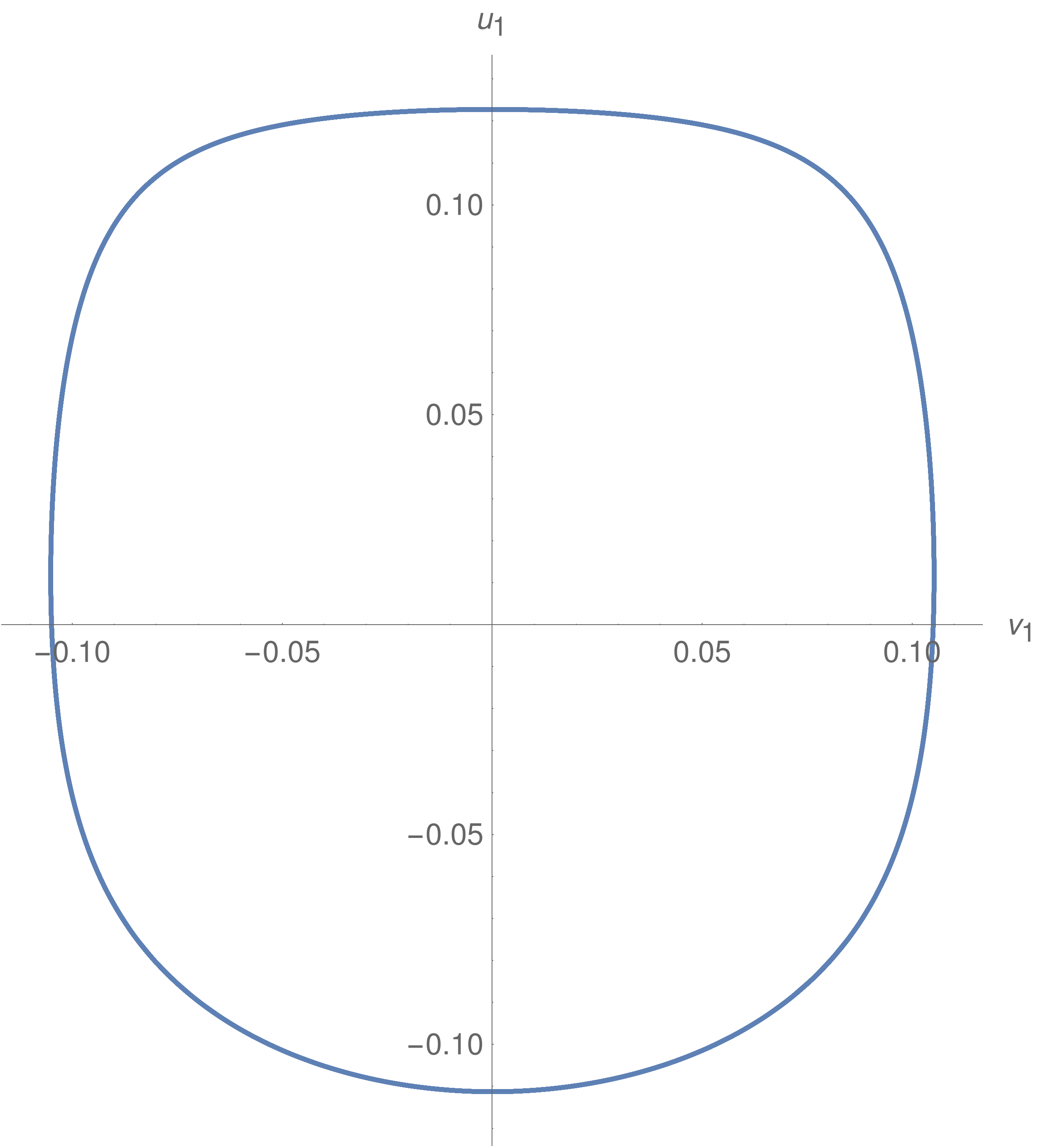}
  \caption{The orbit representing the slow dynamics of the integrable
    approximation $\mathcal{Z}^{(6)}$ in the phase plane endowed with
    coordinates $(v_1,u_1)$.}
  \label{plot-circv1u1}
\end{figure}

\begin{figure}[h]
  \centering
  \includegraphics[scale=0.55]{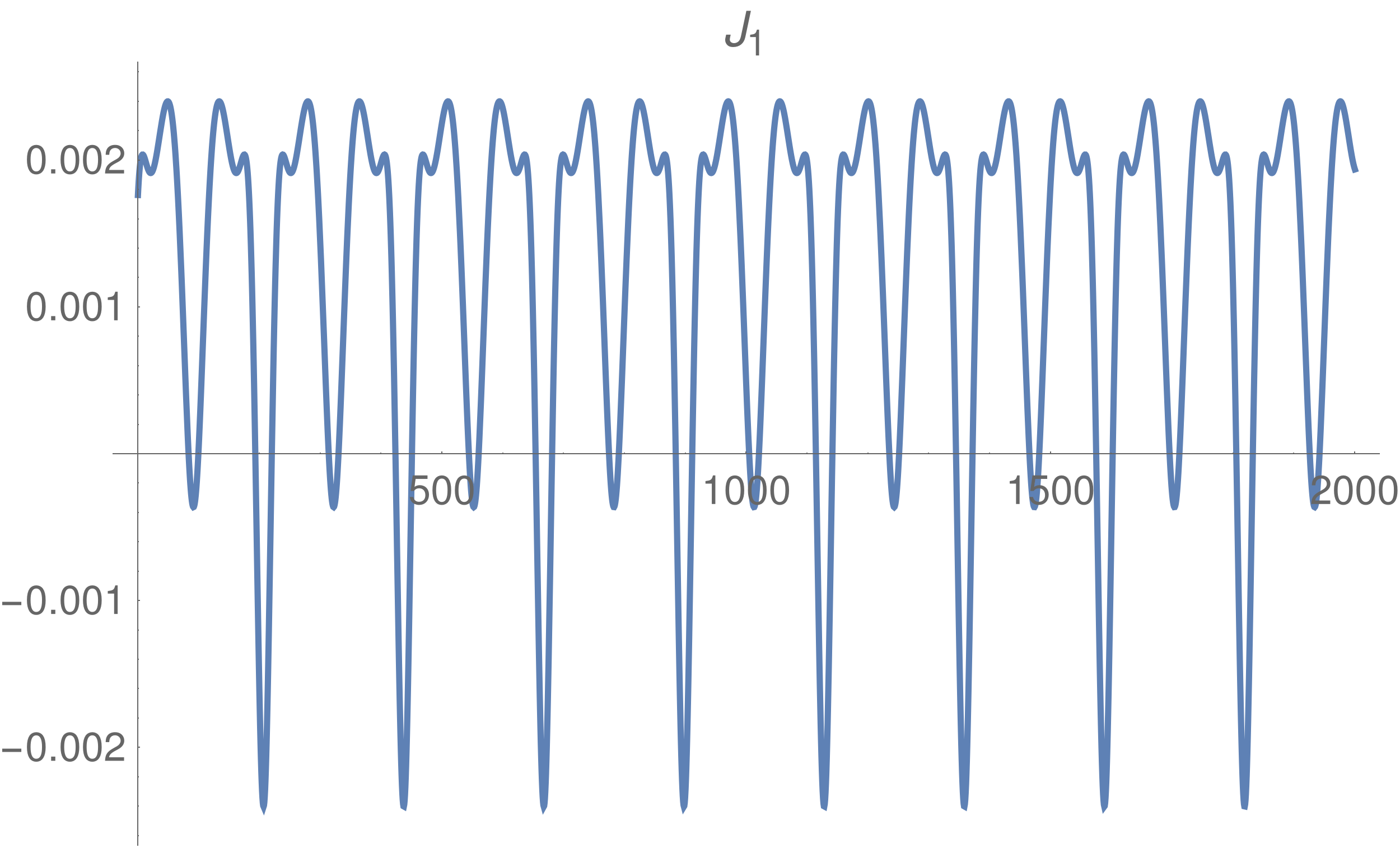}
  \hskip 10 pt
  \includegraphics[scale=0.55]{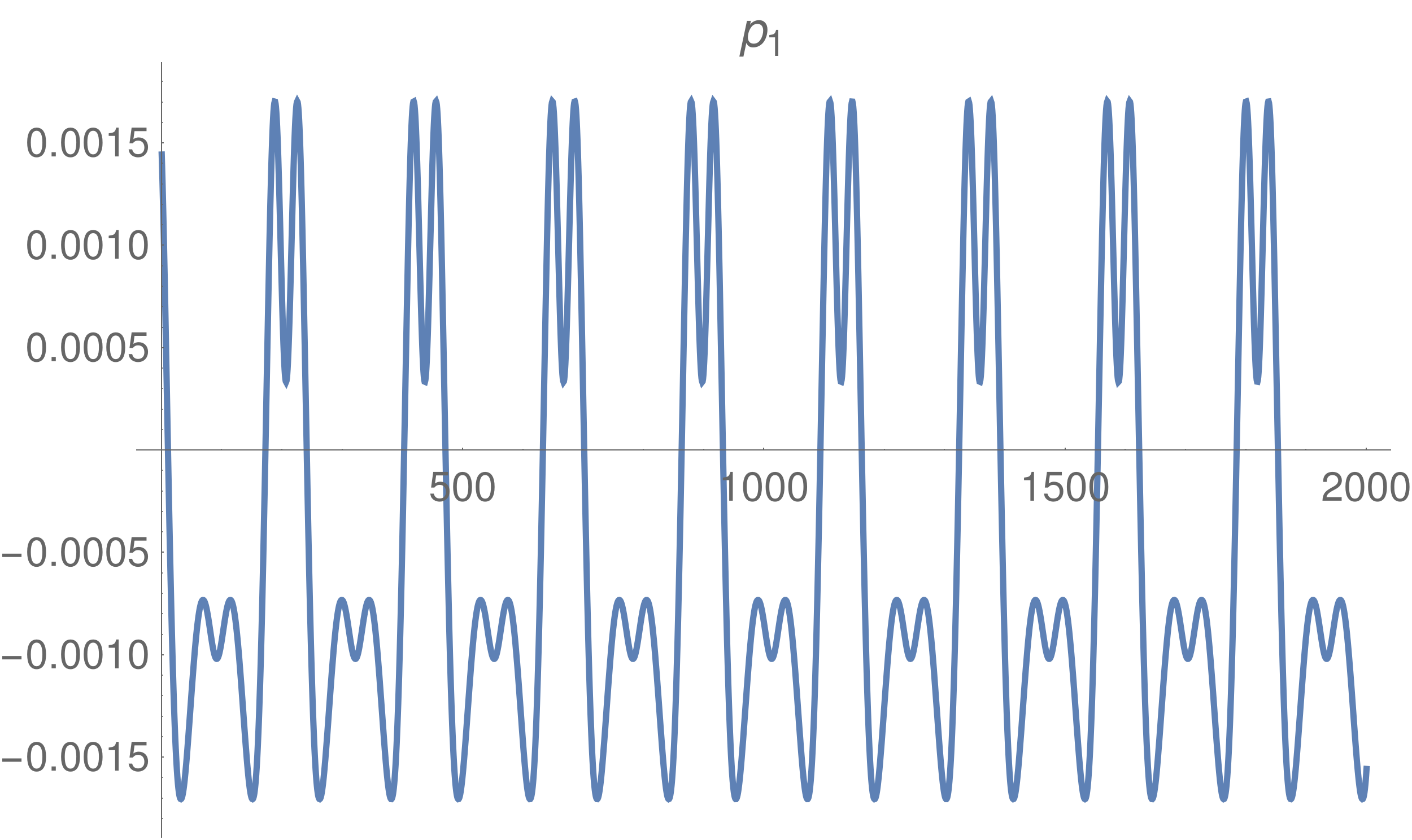}
  \caption{On the left, variation in time of the action
    $J_1=\big(Y_1^2+X_1^2\big)/2$ (with respect to its mid value)
    along the flow induced by the integrable approximation
    $\mathcal{Z}^{(6)}$. On the right, the same plot is made for what
    concerns the action $\big(v_1^2+u_1^2\big)/2$.}
  \label{plot_prima_dopo}
\end{figure}

The new orbit of the slow motion in the variables $(v_1,u_1)$ is
represented in Fig.~\ref{plot-circv1u1}.  Let us remark that this plot
does not represent exactly a circular orbit; this was somehow expected
since we have considered only a limited number of Fourier components
in the computational method we have introduced in the present Section
with the aim of trying to circularize the orbit itself.  However, by
looking at the scales reported on the vertical axes of the two panels
included in Fig.~\ref{plot_prima_dopo}, one can appreciate that the
canonical change of coordinates~\eqref{eq:trasf2morecircular} allows
us to reduce the oscillations of the value of the action involved in
the description of the slow dynamics. In fact, when the plot of the
motion law $t\mapsto\big(Y_1^2(t)+X_1^2(t)\big)/2$ is compared to the
one of $t\mapsto\big(v_1^2(t)+u_1^2(t)\big)/2$, the gain of about 30\%
in the circularization of the orbit is highlighted.  This is enough
for the purpose of obtaining a Kolmogorov normalization algorithm
which is convergent to the normal form related to the desired final
invariant torus.

We can now introduce the action-angle variables that are more
suitably adapted to the integrable approximation, i.e.,
\begin{equation}
  \begin{aligned}
  \label{trasf-con-trasl-iniziale}
  v_1&=\sqrt{2(p_1+p_1^*)}\cos(q_1)\ ,
  \quad &Y_2&=\sqrt{2J_2}\cos(\vartheta_2)\ ,\\
  u_1&=\sqrt{2(p_1+p_1^*)}\sin(q_1)\ , 
  \quad &X_2&=\sqrt{2J_2}\sin(\vartheta_2)\ ,
  \end{aligned}
\end{equation}
where $p_1^*$ is the value of the area enclosed by the orbit that
describes the secular dynamics in the phase plane $(v_1,u_1)$ (or in
the one endowed with coordinates $(Y_1,X_1)$, since canonical
transformations preserve the areas) multiplied by the factor
$1/(2\pi)$. Therefore, we are imposing that the value $p_1^*\,$, which
corresponds to the closed curve $\big\{(p_1,q_1)\,:\ p_1=0\,,\>
q_1\in\toro\big\}$, is equal to the usual definition of the action for
Hamiltonian systems with one degree of freedom (see, e.g., Chap.~3
of~\cite{Giorgilli-Libro-2022}).


\section{Construction of the KAM torus}\label{sec:KAM}

We can now start the construction of the KAM torus for the averaged dynamics of HD60532.  First, we perform a
translation of the fast action and we rename the fast angle, i.e.
\begin{equation}
 \label{trasf-con-trasl-seconda}
 p_2=J_2-J_2^* \ , \qquad q_2=\vartheta_2\ ,
\end{equation}
where $J_2^*$ is the mean value of the action $J_2$.  In the new
action-angle variables $(\bm p,\bm q)$ the Hamiltonian
\eqref{Hamavg-step-r} can be expanded as follows
\begin{equation}
\label{eq:H0}
\vcenter{\openup1\jot\halign{
 \hbox {\hfil $\displaystyle {#}$}
&\hbox {\hfil $\displaystyle {#}$\hfil}
&\hbox {$\displaystyle {#}$\hfil}\cr
 H^{(0)}(\bm p,\bm q) &=
 &E^{(0)}+\bm \omega^{(0)}\cdot \bm p
 +\sum_{s\geq 0}\sum_{\ell\geq 2} f_\ell^{(0,s)}(\bm p,\bm q)
 \cr
 & &+\sum_{s\geq 1}\left( f_0^{(0,s)}(\bm q) + f_1^{(0,s)}(\bm p,\bm q)\right)\ ,
 \cr
}}
\end{equation}
where $f_\ell^{(0,s)}$ is a homogeneous polynomial of degree $\ell$ in
$\bm p$ and a trigonometric polynomial of degree\footnote{More
  generically, the functions $f_\ell^{(0,s)}$ are usually defined as
  trigonometric polynomials of degree $sK$ (for some positive fixed
  value of the parameter $K\in\naturali$) in $\bm q$. We choose to set
  $K=2$, accordingly to what is usually done for quasi-integrable
  Hamiltonian system that are in the vicinity of an elliptic
  equilibrium point as it is in the model we are studying (see,
  e.g.,~\cite{Gio-Loc-San-2017} and recall the discussion in
  Section~\ref{sec:case-study}).} $2s$ in $\bm q$. The first
superscript of the functions $f_\ell^{(0,s)}$ denotes the
normalization step.  Furthermore, $E^{(0)}$ is the constant of the
energy level of $\bm p=0$ when $f_\ell^{(0,s)}=0$ $\,\forall\ \ell$
and $s=0,\,1$.  The goal is to construct the Kolmogorov normal form
\begin{equation}
  \label{KAM-infty}
  H^{(\infty)}(\bm p,\bm q)=\bm \omega^*\cdot\bm p +\Oscr(\|\bm p\|^2)\ ,
\end{equation}
where $\bm \omega^*$ is the angular velocity vector characterizing the
quasi-periodic motion on the invariant (KAM) torus corresponding to
$\bm p=\bm 0$. In other words, the Kolmogorov normalization algorithm
is designed in such a way to remove the terms appearing in the second
row of formula~\eqref{eq:H0} by a sequence of canonical
transformations.  Here, it is convenient to adopt a different version
of the classical Kolmogorov normalization algorithm, which is slightly
modified in such a way to not keep fixed the angular velocity vector
$\bm \omega^{(r)}$, that is defined at the $r$-th step of the
procedure and corresponds to the quasi-periodic approximation of the
motion on the final sought KAM torus. We basically follow the approach
described in~\cite{Loc-Car-San-Vol-2022}, where the normalization procedure introduced by Kolmogorov is adapted in such a way to skip the small translation of the actions performed at every step of that algorithm. This modification allows to make the computational procedure more stable; such an improvement can play a crucial role when the action--frequency map is (close to be) degenerate (see \cite{Gab-Jor-Loc-2005}). Moreover, in order to improve its efficiency, this small adaptation of the Kolmogorov normalization algorithm has to be formulated so as to suitably determine the preliminary translation in \eqref{trasf-con-trasl-iniziale}. All this computational procedure is summarized in the
following in order to make our discussion rather
self-consistent.

As in Section~\ref{sec:average}, it is convenient to introduce suitable
classes of functions; here, we are going to say that $g\in\Pgot_{\ell,sK}$
if its Taylor-Fourier expansion writes as
\begin{equation}
  \label{eq:Taylor-Fourier-per-generica-g}
  g(\bm p, \bm q) =
  \sum_{{\scriptstyle{\bm j\in\naturali^{n}}}\atop{\scriptstyle{| \bm j|=\ell}}}
  \sum_{{\scriptstyle{{ \bm k\in\interi^{n}}}\atop{\scriptstyle{| \bm k|\le sK}}}} 
  c_{\bm j,\bm k}\,\bm p^{\bm j}\exp(\imunit{\bm k}\cdot{\bm q}) \ ,
\end{equation}
for some fixed values of the non-negative integer parameters
$\ell$, $s$ and $K$. The following statement allows us to describe the
behaviour of such a class of functions with respect to the Poisson
brackets.
\begin{lemma}
  \label{lem:classi-funzioni-KAM}
  Let us consider two generic functions $g\in\Pgot_{\ell,sK}$ and
  $h\in\Pgot_{m,rK}\,$, where $K$ is a fixed positive integer
  number. Then, the following inclusion property holds
  true:
  $$
  \big\{g,h\big\} = \lie{h}\,g \in \Pscr_{\ell+m-1,(r+s)K}
  \quad
  \ \forall\>\ell,\,m,\,r,\,s\in\naturali
  \ {\rm with}\ \ell+m\ge 1\ ,
  $$
  while $\big\{g,h\big\} = 0$ when $\ell=m=0$.
\end{lemma}

Let us imagine to have already performed $r-1$ normalization steps by
using, once again, the Lie series formalism; then, we have to
deal with an Hamiltonian of the following type:
\begin{equation}
\label{eq:Hr-1}
\vcenter{\openup1\jot\halign{
 \hbox {\hfil $\displaystyle {#}$}
&\hbox {\hfil $\displaystyle {#}$\hfil}
&\hbox {$\displaystyle {#}$\hfil}\cr
 H^{(r-1)}(\bm p,\bm q) &=
 &E^{(r-1)}+\bm \omega^{(r-1)}\cdot \bm p
 +\sum_{s\geq 0}\sum_{\ell\geq 2} f_\ell^{(r-1,s)}(\bm p,\bm q)
 \cr
 & &+\sum_{s\geq r}\left( f_0^{(r-1,s)}(\bm q) + f_1^{(r-1,s)}(\bm p,\bm q)\right)
 \ ,
 \cr
}}
\end{equation}
where $f_\ell^{(r-1,s)}\in\Pgot_{\ell,2s}$
$\forall\ \ell,\,s\,\in\naturali$, while $E^{(r-1)}\in\reali$. Let us
remark that the expansion of $H^{(0)}$, which is reported
in~\eqref{eq:H0}, agrees with the more general one, that is written
just above in~\eqref{eq:Hr-1}, in the case with $r=1$. The Kolmogorov
normalization algorithm at step $r$ is aimed to remove the main
perturbing terms (that are the functions $f_0^{(r-1,r)}$ and
$f_1^{(r-1,r)}$), which are independent of and linear in the actions,
respectively.

In order to perform the $r$-th normalization step, first we need to
determine the generating function $\chi_0^{(r)}$ in such a way to
solve the following homological equation:
\begin{equation*}
  \lie{\chi_0^{(r)}}\left(\bm \omega^{(r-1)}\cdot \bm p\right) +f_0^{(r-1,r)}=
  \langle f_0^{(r-1,r)}\rangle_{\bm q}\ .
\end{equation*}
As a matter of fact,
$\langle f_0^{(r-1,r)}\rangle_{\bm q}\in\Pgot_{0,0}$ is nothing but a
constant term. Therefore, it can be added to $E^{(r-1)}$, in order
to update the energy level, whose new value is denoted with $E^{(r)}$.
By considering the Taylor-Fourier expansion of the perturbing term we
aim to remove, i.e.,
\begin{equation*}
  f_0^{(r-1,r)}(\bm q)=
  \sum_{0<|\bm k|\leq 2r} c_{\bm 0,\bm k}^{(r-1,r)}\exp(\imunit \bm k\cdot \bm q)\ ,
\end{equation*}
we obtain the following expression for the generating function:
\begin{equation*}
  \chi_0^{(r)}(\bm q) = \sum_{0<|\bm k|\leq 2r}
  \frac{c_{\bm 0,\bm k}^{(r-1,r)}}{\imunit  \bm k\cdot\bm \omega^{(r-1)}}
  \exp(\imunit \bm k\cdot \bm q)\ . 
\end{equation*}
Let us remark that the homological equation can be solved provided
that the following non-resonance condition holds true
\begin{equation}
  \label{non-resonance}
  \bm k\cdot\bm \omega^{(r-1)}\neq 0\ , \qquad
  \forall\, \bm k\in\interi^2\setminus\{0\}\ ,
  \quad \text{with}\,\, |\bm k|\leq 2r\ .
\end{equation}

We then introduce the transformed Hamiltonian
$\hat{H}^{(r)}=\exp\left(\lie{\chi_0^{(r)}}\right)H^{(r-1)}$,
whose expansion
\begin{equation}
\label{eq:hatHr}
\vcenter{\openup1\jot\halign{
 \hbox {\hfil $\displaystyle {#}$}
&\hbox {\hfil $\displaystyle {#}$\hfil}
&\hbox {$\displaystyle {#}$\hfil}\cr
 {\hat H^{(r)}}(\bm p,\bm q) &=
 &{\hat E}^{(r)}+\bm \omega^{(r-1)}\cdot \bm p
 +\sum_{s\geq 0}\sum_{\ell\geq 2} {\hat f}_\ell^{(r,s)}(\bm p,\bm q)
 \cr
 & &+\sum_{s\geq r}\left( {\hat f}_0^{(r,s)}(\bm q) +
 {\hat f}_1^{(r,s)}(\bm p,\bm q)\right)
 \cr
}}
\end{equation}
is such that the new Hamiltonian terms $\hat{f}_\ell^{(r,s)}$ are defined
so that
\begin{equation*}
  \begin{aligned}
    \hat{f}_{0}^{(r,r)} &= 0 \ , \\
    \hat{f}_{\ell}^{(r,s)}&= \sum_{j=0}^{\lfloor s/r \rfloor} \frac{1}{j!}
    \lie{\chi^{(r)}_{0}}^{j} f^{(r-1,s-jr)}_{\ell+j} \ ,
    \qquad\qquad\hbox{for }{\vtop{\hbox{$\ell=0,\ s\neq r$\ ,}
        \vskip-2pt\hbox{\hskip-14pt or $\ell\neq0 \ s\geq 0$ \ .}}}
  \end{aligned}
\end{equation*}
By applying repeatedly Lemma~\ref{lem:classi-funzioni-KAM}, one
can easily verify that $\hat{f}_{\ell}^{(r,s)}\in\Pgot_{\ell,2s}$
$\forall\ \ell,\,s$.

The second generating function $\chi_1^{(r)}$ is determined by solving
the following homological equation:
\begin{equation}
  \label{second-homological}
  \lie{\chi_1^{(r)}}\left(\bm \omega^{(r-1)}\cdot \bm p \right)+
  \hat{f}_1^{(r,r)}=\langle \hat{f}_1^{(r,r)}\rangle_{\bm q}\ .
\end{equation}
The term $\langle\hat{f}_1^{(r,r)}\rangle_{\bm q}\in\Pgot_{1,0}$. This
means that it is not dependent on the angles $\bm q$ and is linear in
the actions $\bm p$; thus, it gives a contribution to the definition
of the value of the angular velocity vector $\bm \omega^{(r)}$, which
in principle should converge to its limit $\bm \omega^{*}$ (if the
normalization algorithm is convergent) and is defined so that
\begin{equation*}
  \bm \omega^{(r)}\cdot\bm p=
  \bm \omega^{(r-1)}\cdot\bm p +\langle \hat{f}_1^{(r,r)}\rangle_{\bm q}\ .
\end{equation*}

By considering the following Taylor-Fourier expansion of the new
perturbing term we aim to remove, i.e.
\begin{equation*}
  \hat{f}_1^{(r,r)}(\bm q) - \langle \hat{f}_1^{(r,r)}\rangle_{\bm q} =
  \sum_{|\bm \ell|=1}\,
  \sum_{0<|\bm k|\leq 2r} \hat{c}_{\bm \ell,\bm k}^{(r,r)}
  {\bm p}^{\bm \ell}\exp(\imunit \bm k\cdot \bm q)\ ,
\end{equation*}
then we easily determine the new generating function as
\begin{equation*}
  \chi_1^{(r)}(\bm q)= \sum_{|\bm \ell|=1}\,
  \sum_{0<|\bm k|\leq 2r}
  \frac{\hat{c}_{\bm \ell,\bm k}^{(r,r)}}{\imunit \bm k\cdot\bm \omega^{(r-1)}}
  {\bm p}^{\bm \ell}\exp(\imunit \bm k\cdot \bm q)\ . 
\end{equation*}
Once again, the homological equation can be solved provided that the
frequencies satisfy the non resonance condition~\eqref{non-resonance}.
The new Hamiltonian is defined as
$H^{(r)}=\exp\left(\lie{\chi_1^{(r)}}\right)\hat{H}^{(r-1)}$. Its
expansion is completely analogous to the one reported
in~\eqref{eq:Hr-1}. Moreover,
the terms $f_{\ell}^{(r,s)}\in\Pgot_{\ell,2s}$ appearing in the
expansion of $H^{(r)}$ are defined in the following way:
\begin{equation*}
 \begin{aligned}
f_{1}^{(r,r)} &= 0 \ , \\
f_{1}^{(r,ir)} &=
\frac{i-1}{i!}\lie{\chi^{(r)}_{1}}^{i-1}\hat{f}^{(r,r)}_{1}+
\sum_{j=0}^{i-2} \frac{1}{j!}
\lie{\chi^{(r)}_{1}}^{j} \hat{f}^{(r,(i-j)r)}_{1} \ ,
\qquad\qquad\hbox{for }{\vtop{\hbox{$ i\geq 2$\ ,}}}\\ 
f_{\ell}^{(r,s)} &=
\sum_{j=0}^{\lfloor s/r \rfloor} \frac{1}{j!} \lie{\chi^{(r)}_{1}}^{j}
\hat{f}^{(r,s-jr)}_{\ell} \ ,
\qquad\qquad\hbox{for }{\vtop{\hbox{$\ell=1,\ s\neq ir$\ ,}
    \vskip-2pt\hbox{\hskip-14pt or $\ell\neq1, \ s\geq 0$ \ ,}}}
\end{aligned}
\end{equation*}
where we have exploited the second homological
equation~\eqref{second-homological}.

From a practical point of view, we can iterate the algorithm only up
to a finite number of steps, say, $\overline{r}$. This allows us to
determine
\begin{equation}
  \label{KAM-normal-form}
  H^{(\overline{r})}(\bm p,\bm q)=E^{(\overline{r})}
  +\bm \omega^{(\overline{r})}\cdot \bm p
  +\sum_{s\geq 0}\sum_{\ell\geq 2} f_\ell^{(\overline{r},s)}(\bm p,\bm q)
  +\sum_{s\geq \overline{r}+1}
  \left( f_0^{(\overline{r},s)}(\bm q) + f_1^{(\overline{r},s)}(\bm p,\bm q)\right)
  \ .
\end{equation}
Hence, we obtain an approximation of the final invariant torus which
is characterized by an angular velocity vector
$\bm\omega^{(\overline{r})}$.  If the value (say) $I_1^*$ of the
initial shift on the first action, which has been preliminarly fixed
equal to $p_1^*$ in formula~\eqref{trasf-con-trasl-iniziale}, is accurate
enough, then the slow frequency $\omega^{(\overline{r})}_1$ is close
to the one we are aiming at, which is numerically determined by
applying the frequency analysis method, namely
$\omega^{(\overline{r})}_1\simeq \omega_1^*$.  We then calibrate the
initial translation of the first action $I_1^*=p_1^*$ by means of a
Newton method.  The goal is to solve the implicit equation
$\omega_1(\tilde{I}_1)=\omega_1^*$ with respect to the initial
shift $\tilde{I}_1$.  The value $\tilde{I}_1$ is iteratively
computed using the formula
\begin{equation*}
  \tilde{I}_1^{(n)}=\tilde{I}_1^{(n-1)}+
  \frac{\omega_1^*-\omega_1^{(\overline{r})}\left(\tilde{I}_1^{(n-1)}\right)}
       {\omega_1^\prime\left(\tilde{I}_1^{(n-1)}\right)}\ ,
\end{equation*}
where $\tilde{I}_1^{(0)}=I_1^*$ and the value of the derivative
$\omega_1^\prime\big(\tilde{I}_1^{(n-1)}\big)$ is numerically
approximated by using the finite difference method.  Let us recall
that, after having performed the average with respect to the fast
angle of libration as it has been described in the previous section,
we are mainly focusing on the study of the secular dynamics.  In
addition, for what concerns the (relatively) faster frequency
$\omega_2^{(\overline{r})}/(2\pi)$ we automatically have a good enough
approximation of both the frequencies of the averaged Hamiltonian, as
it can be appreciated looking at the comparison between the
semi-analytic solutions showed in Fig.~\ref{fig:media-KAM}, which will
be widely commented in the next subsection.

By supposing to iterate the normalization algorithm ad infinitum, one
would get the Hamiltonian \eqref{KAM-infty}, which admits the
invariant torus $\bm p=0$ with frequency $\bm\omega^*$.
\begin{figure}[]
  \centering
  \includegraphics[scale=0.50]{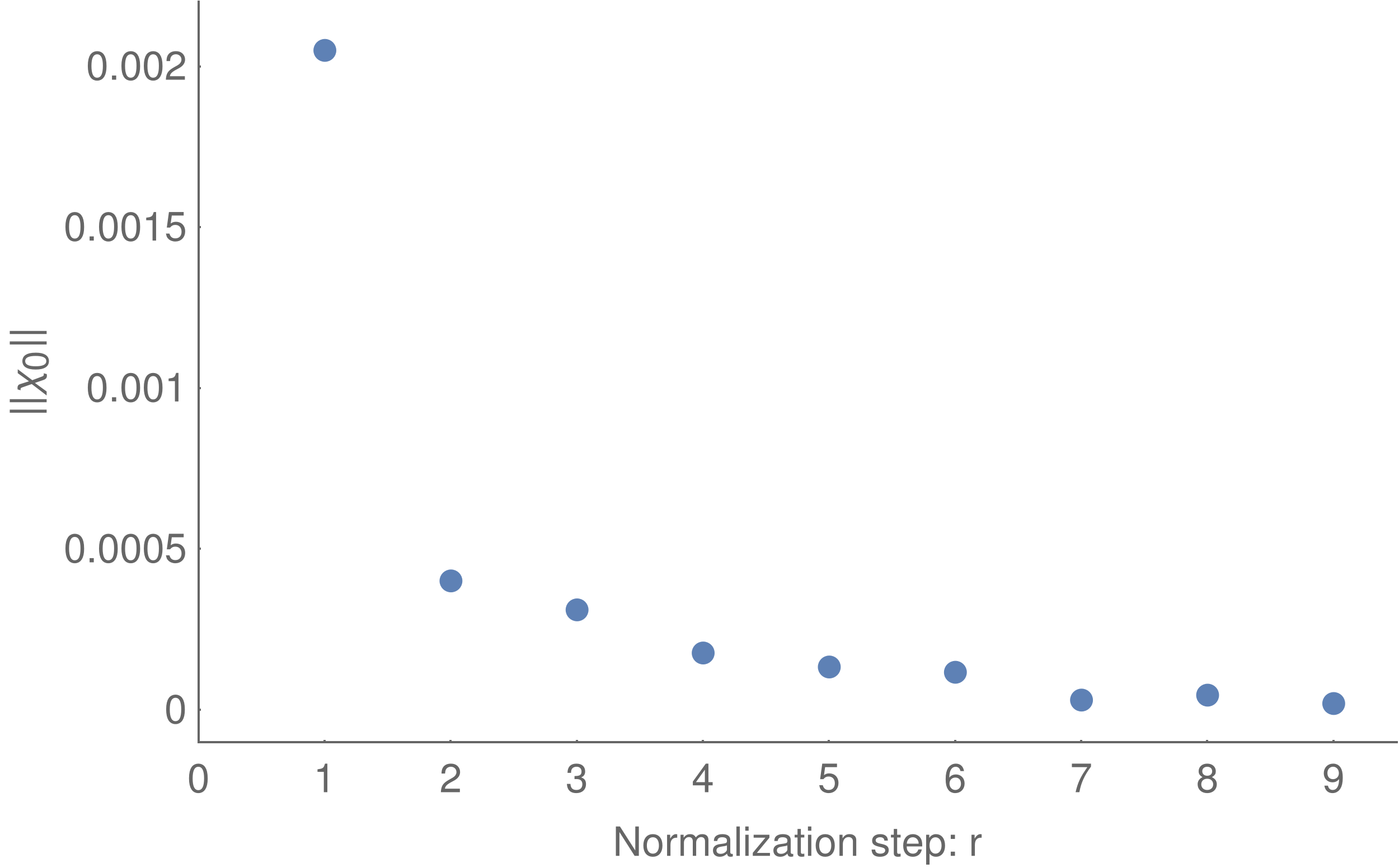}
  \hskip 10pt	\includegraphics[scale=0.50]{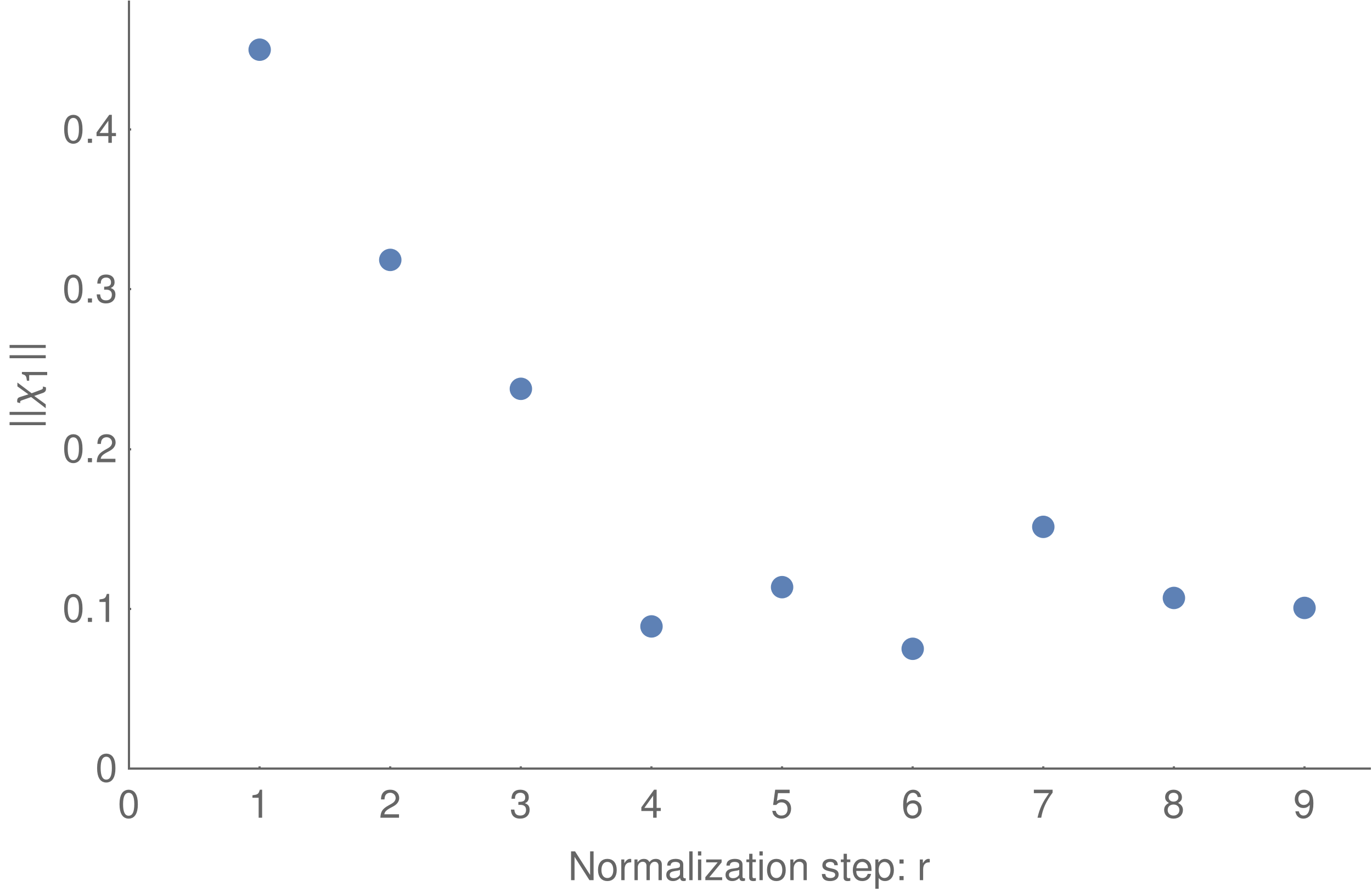}
  \caption{Norms of the generating functions $\chi_0^{(r)}$ and $\chi_1^{(r)}$ as they
    are determined by the Kolmogorov algorithm up to the $9$-th
    normalization step.}
  \label{norms-gen-funct}
\end{figure}
From a practical point of view, we are able to explicitly iterate the
algorithm only up to a finite normalization step $\overline{r}$ and we
can numerically check the convergence of the procedure by controlling
the decrease of the norms of the generating functions. Hereafter, we
define the norm of any generic function $g\in\Pgot_{\ell,sK}$ as
$$
\|g(\bm p, \bm q)\| =
\sum_{{\scriptstyle{\bm j\in\naturali^{n}}}\atop{\scriptstyle{| \bm j|=\ell}}}
\sum_{{\scriptstyle{{ \bm k\in\interi^{n}}}\atop{\scriptstyle{| \bm k|\le sK}}}} 
\left|c_{\bm j,\bm k}\right|\ ,
$$
$\forall\ \ell,\,s,\,K\,\in\naturali$, where the Taylor-Fourier
expansion of $g$ is written
in~\eqref{eq:Taylor-Fourier-per-generica-g}.  The behavior of $\|\chi_0^{(r)}\|$ and
$\|\chi_1^{(r)}\|$ for values of the normalization step $r$ up to~$9$
are reported in Fig.~\ref{norms-gen-funct}.

\subsection{Comparison between two different kinds of semi-analytic solutions}\label{second-comparison}

\begin{figure}
  \centering
  \includegraphics[scale=0.50]{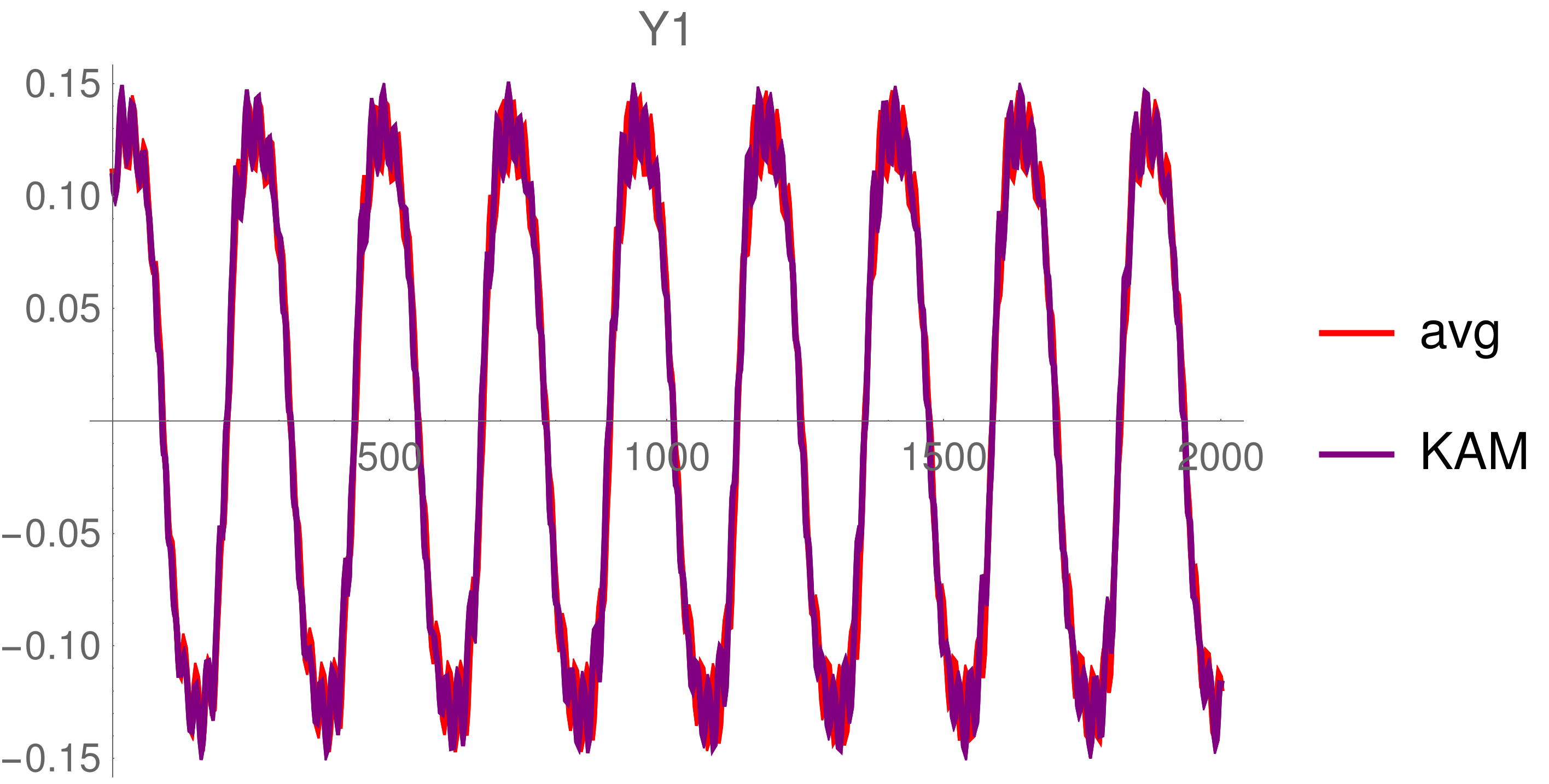}
  \hskip 10pt
  \includegraphics[scale=0.50]{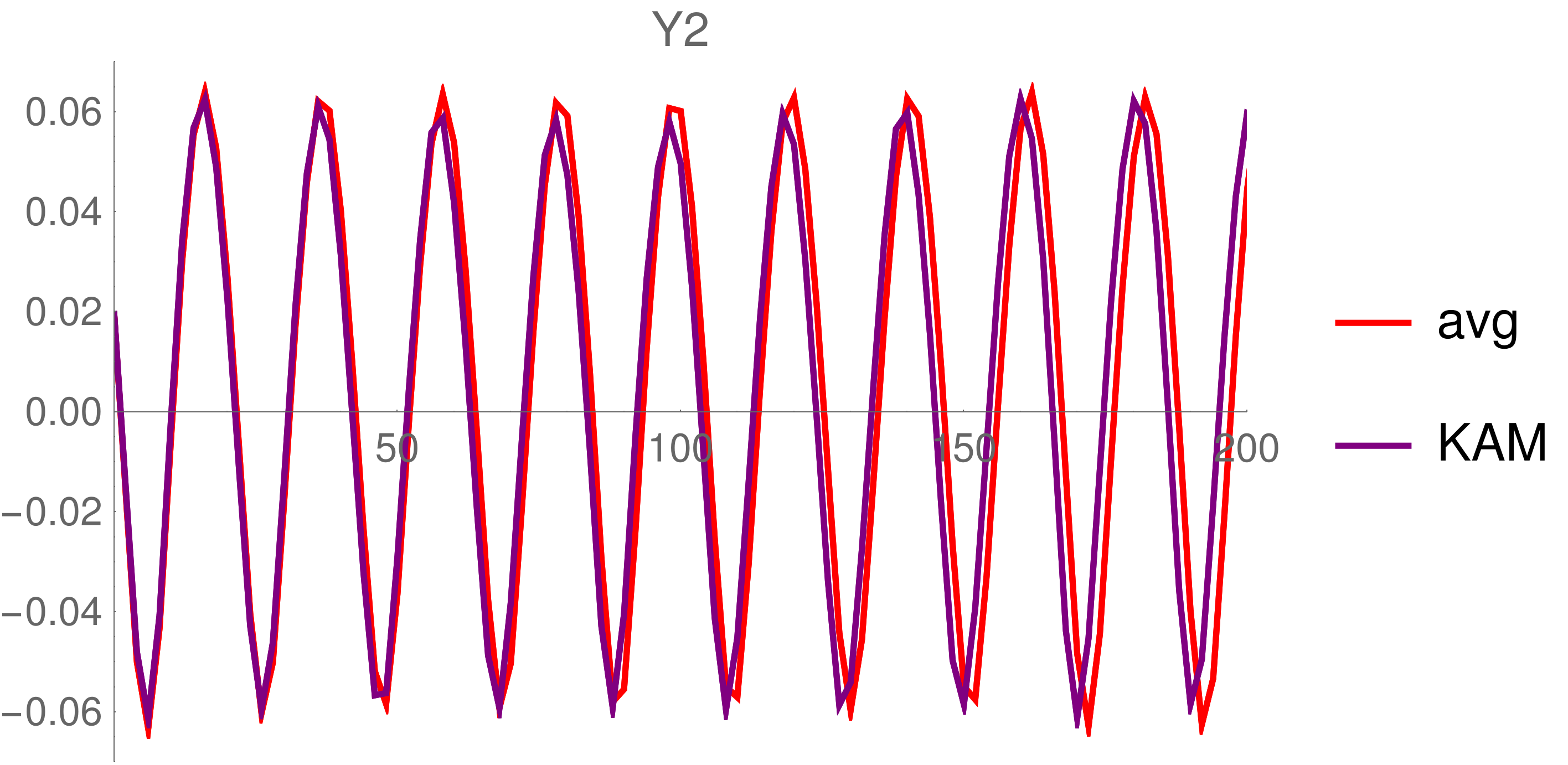}
  \includegraphics[scale=0.50]{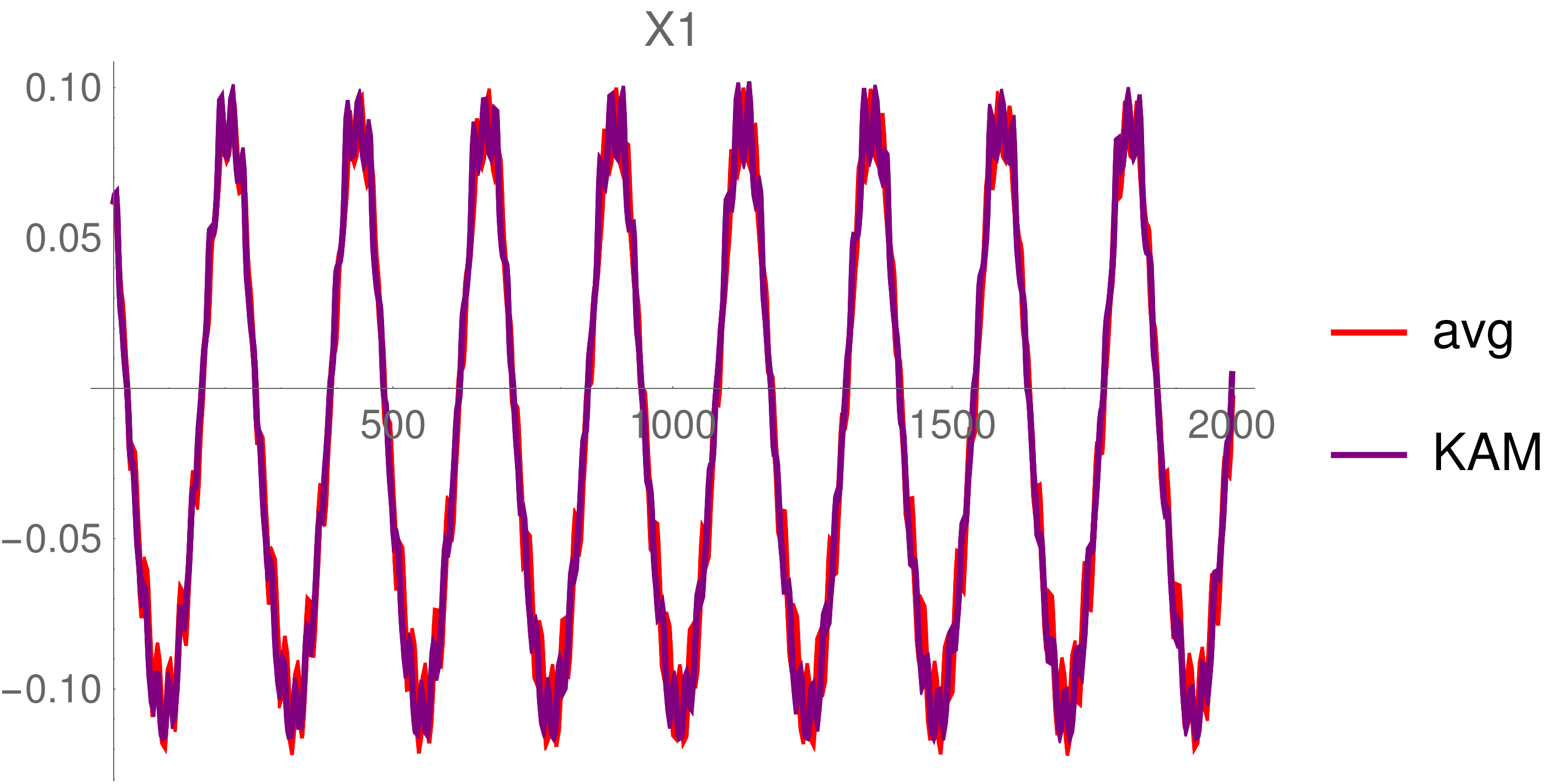}
  \hskip 10pt
  \includegraphics[scale=0.50]{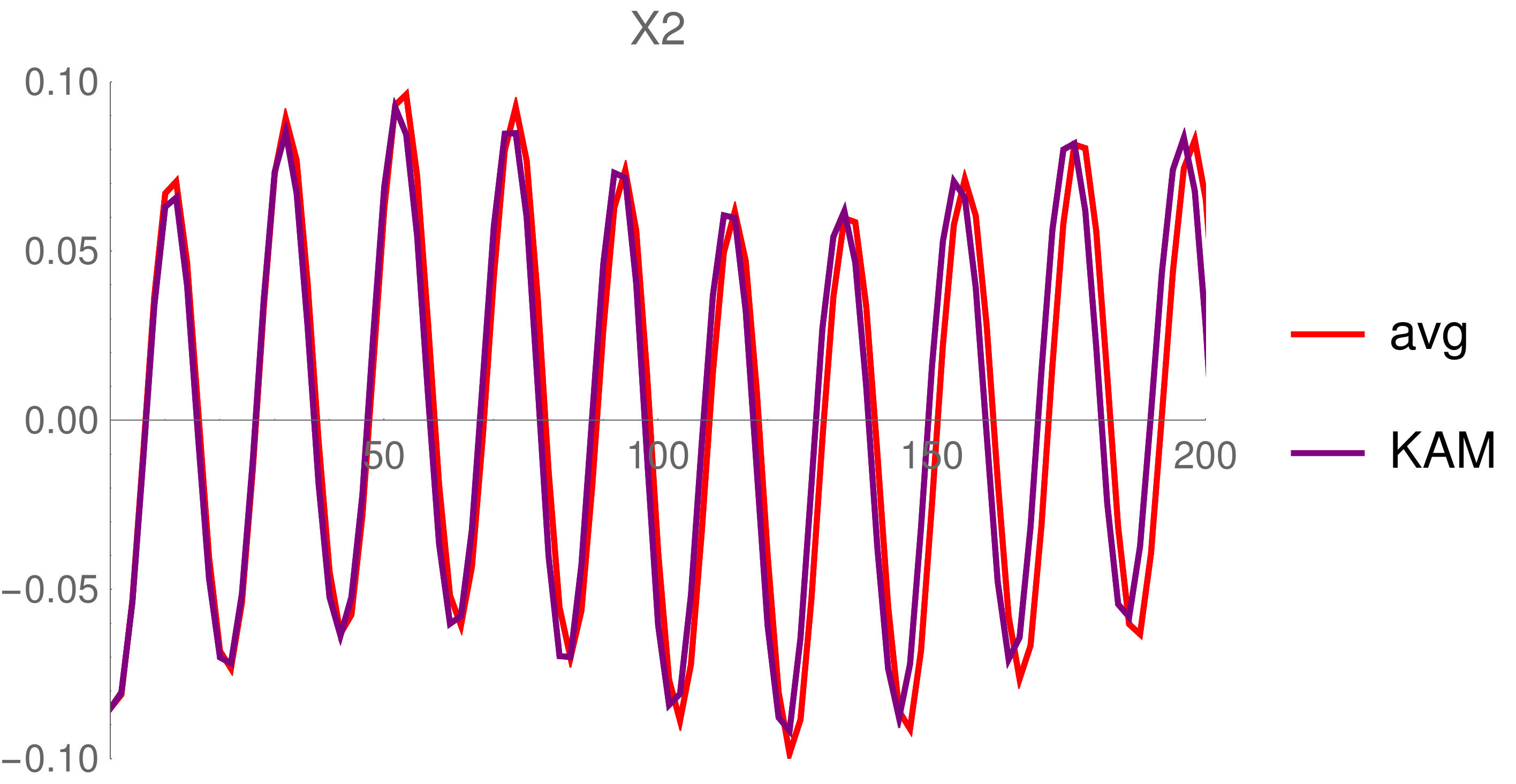}
  \caption{Evolutions with respect to time [yr] of the secular
    variables (reported in the panels on the left) and the relatively
    faster ones (in the right panels) in our model of the MMR
    librational dynamics of HD60532. They are given by the
    semi-analytic solutions of the averaged Hamiltonian up to order
    $6$ (red curves) and the Hamiltonian that is produced after $5$
    steps of the Kolmogorov normalization algorithm (purple curves).}
  \label{fig:media-KAM}
\end{figure}

In this subsection we check the accuracy of the
Hamiltonian~\eqref{KAM-normal-form} in Kolmogorov normal form up to a
finite order $\overline{r}$ in describing the motion of the averaged
integrable Hamiltonian $\mathcal{Z}^{(\tilde r)}$ up to order $\tilde
r$.  For what concerns our model of the librational dynamics of the
extrasolar system HD60532, we consider the Hamiltonian $H^{(5)}$,
expanded as in \eqref{KAM-normal-form} and truncated up to degree $2$
in the actions and to trigonometrical degree $12$ in the angles.  The
aim is to make a comparison with the solution associated to the
averaged integrable Hamiltonian $\mathcal{Z}^{(6)}$, computed in
Subsection~\ref{sec:first-comparison}.  The semi-analytic solution of
the equations of motion which is related to the Hamiltonian $H^{(5)}$
can be obtained with a procedure similar to the one previously
described and represented in~\eqref{semi-analytical_scheme}.
Moreover, as we can see in Fig.~\ref{fig:media-KAM}, we compare the
motion laws induced by two different Hamiltonians by considering in
both cases the cartesian variables $Y_j=\sqrt{2J_j}\cos(\vartheta_j)$
and $X_j=\sqrt{2J_j}\sin(\vartheta_j)$, for $j=1,\,2$, that were
adopted as canonical coordinates before starting the averaging
procedure which constructs the resonant Birkhoff normal form. In more
detail, we can determine the expansions of all the canonical
transformations introduced in Sections~\ref{sec:average},
\ref{sec:action-angle} and~\ref{sec:KAM} with the aim of constructing
a Hamiltonian in Kolmogorov normal form up to order $5$. Let us denote
with the symbol $\mathcal{K}^{(5)}$ the composition of the canonical
transformations introduced by the Kolmogorov algorithm (described in
the previous Subsection) up to the $5$-th normalization step, i.e.,
\begin{equation}
  \mathcal{K}^{(5)}(\bm p,\bm q)=
  \exp\lie{\chi_1^{(5)}}\,\circ\,\exp\lie{\chi_0^{(5)}}
  \,\circ\,\ldots\,\circ
  \exp\lie{\chi_1^{(1)}}\,\circ\,\exp\lie{\chi_0^{(1)}}
  \,(\bm p,\bm q)\ .
\label{eq:trasf_Kolm_norm}
\end{equation}
Therefore, $\forall\ t\in\reali$, we can compute the values of the
canonical variables $(\bm Y(t),\bm X(t))$ corresponding to the $(\bm
p(t),\bm q(t))=(0,\bm \omega^{(5)} t+\bm q(0))$, which describe the
quasi-periodic motion of the final KAM torus as it is {\it
  approximately} reproduced by the Kolmogorov normalization algorithm,
when it is iterated up to the $5$-th step. This computation is
performed according to the following scheme:
\begin{equation}
\begin{tikzcd}[row sep=5em, column sep=10em,every label/.append style = {font = \normalsize}]
  \big(\bm Y(0),\bm X(0)\big) \arrow[r, "\left(\Ascr\circ\mathcal{C}^{(5)}\circ\mathcal{T}_{\tilde{I}_1}\circ\mathcal{K}^{(5)}\right)^{-1}",shorten <=1em,shorten >=1em] & \big(\bm 0,\bm q(0)\big) \arrow[d, "\Phi_{\bm \omega^{(5)} \cdot \bm p}^t",shorten <=0.5em,shorten >=0.5em] \\
  \big(\bm Y(t),\bm X(t)\big) \arrow[r,leftarrow,"\Ascr\circ\mathcal{C}^{(5)}\circ\mathcal{T}_{\tilde{I}_1}\circ\mathcal{K}^{(5)}",shorten <=0.5em,shorten >=0.5em] & \big(\bm p(t)=\bm 0,\bm q(t)=\bm \omega^{(5)} t+\bm q(0)\big)
\end{tikzcd}
\label{semi-analytical_scheme_KAM}
\end{equation}
Here a few further explanations are in order. We denote with $(\bm
J,\bm \vartheta)=\mathcal{T}_{\tilde{I}_1}(\bm p,\bm q)$ the canonical
transformation that is obtained by making the composition of all the
canonical transformations described in Section~\ref{sec:action-angle}
and in formula~\eqref{trasf-con-trasl-seconda}; moreover, one has to
take care of slightly modifying~\eqref{trasf-con-trasl-iniziale} in
such a way to replace $p_1^*$ with the value of $\tilde{I}_1$ (i.e.,
the solution of equation $\omega_1(\tilde{I}_1)=\omega_1^*$
numerically obtained by applying the Newton method). In the
scheme~\eqref{semi-analytical_scheme_KAM} we have also decided to
consider $\mathcal{C}^{(5)}$ instead of $\mathcal{C}^{(6)}$, because
otherwise with the adopted rules of truncation the Hamiltonian would
be integrable already before the Kolmogorov normalization; this would
make trivial the application of such an algorithm. Finally, let us
recall that $H^{(5)}(\bm p,\bm q)\simeq\bm\omega^{(5)}\cdot\bm p
+\Oscr(\|\bm p\|^2)$ and $\omega^{(5)}\simeq\omega^*$; this allows us
to put the flow of $\bm \omega^{(5)} \cdot \bm p$ in order to
approximate (in the semi-analytical scheme above) the solution of the
equations of motion related to $H^{(5)}$ and with initial conditions
$\bm p(0)=\bm 0$.

The plots in Fig.\,\ref{fig:media-KAM} show an excellent
superimposition between the two solutions, with respect to both the
amplitudes and the frequencies. This makes evident the effectiveness
of our computational algorithm. Let us also recall that the initial
conditions $(\bm Y(0),\bm X(0))$ are the ones compatible with the
observations.

\subsection{Computer-assisted proof}

By looking at the plots in Fig.~\ref{norms-gen-funct}, it can be
noticed that the decrease of the norms of the generating functions, in
particular for what concerns the finite sequence of the second
generating function $\chi_1^{(r)}$, is not so regular and the
convergence of the algorithm looks doubtful.  In order to rigorously
prove that the KAM algorithm is convergent, we adopt a rigorous
approach based on a computer-assisted proof. For this purpose, we
follow the method which has been described in~\cite{CelGL-2000} and
further developed in~\cite{ValLoc-2021}, where a publicly available
software package\footnote{That software package can be freely
  downloaded from the web address {\url{http://dx.doi.org/10.17632/jdx22ysh2s.1}}\label{foot-CAP4KAM}} is provided as
supplementary material. Such a package is designed for doing just this
kind of computer-assisted proof for Hamiltonian systems having two
degrees of freedom. In order to use this software so as to apply it to
the problem under consideration, it is just matter to prepare some
input files, which basically describe the starting Hamiltonian; in
principle, this can allow us to prove the existence of the KAM torus
we are aiming at, if the corresponding Kolmogorov normal form is close
enough to such an initial Hamiltonian. More precisely, we consider
$H^{(5)}$ as the starting Hamiltonian. It is fully determined at the
end of the application of the Newton method, which has been described
in the previous Section; in terms of a single mathematical formula, it
can be written as
$$
H\circ\Ascr\circ\mathcal{C}^{(5)}\circ
\mathcal{T}_{\tilde{I}_1}\circ\mathcal{K}^{(5)}
$$
where $H$, $\Ascr$, $\mathcal{C}^{(5)}$,
$\mathcal{T}_{\tilde{I}_1}$ and $\mathcal{K}^{(5)}$ are defined
in~\eqref{Ham-diag}, \eqref{action-angle-coord},
\eqref{eq:trasf_res_Birkh}, just below
formula~\eqref{semi-analytical_scheme_KAM}
and~\eqref{eq:trasf_Kolm_norm}, respectively.  Moreover, the expansion
of $H^{(5)}$ can be written as in~\eqref{KAM-normal-form} and is
truncated up to degree $2$ in the actions and to trigonometrical
degree $12$ in the angles, while the expansion~\eqref{Hamavg-step-r}
of the intermediate Hamiltonian
$\Hscr^{(5)}=H\circ\Ascr\circ\mathcal{C}^{(5)}$ has been preliminarly
truncated so as to exclude the sum of terms $\sum_{\ell>6}
h_\ell^{(r)}(\bm J,\bm \vartheta)=\Oscr\big(\|\bm J\|^{7/2}\big)$.
Therefore, $H^{(5)}$ is not in Kolmogorov normal form because of a few
(small) Hamiltonian terms that are either dependent on the angles $\bm
q$ only or linearly dependent on the actions $\bm p$.

During the initial stage of the computer-assisted proof, a first code
explicitly performs a (possibly large) number $R_{\rm I}$ of
normalization steps of a classical formulation of the Kolmogorov
algorithm, which includes also small translations of the actions that
aim at keeping fixed the desired angular velocity vector of the
quasi-periodic motion on the final torus, i.e.,
$\bm\omega^*$. Afterwards, the size of the perturbation is further
reduced (although in a less efficient way) by another code which just
iterates the estimates of the norms of the terms of order $r$ with
$R_{\rm I}<r\leq R_{\rm II}$. In the case of the model we are
studying, in order to achieve the convergence of the algorithm, we
have found convenient to set $R_{\rm I}=200$ and $R_{\rm II}=20\,000$.
The whole computer-assisted proof\footnote{The
    software package allowing to perform the computer-assisted proof
    of theorem~\ref{CAP} is available at {\tt
      https://www.mat.uniroma2.it/{\textasciitilde}locatell/CAPs/CAP4KAM-HD60532.zip}\\ As
    a matter of fact, the codes included in this software package are
    exactly the same as the ones which can be downloaded from the
    website mentioned in
    footnote${{\ref{foot-CAP4KAM}}\atop{\phantom{1}}}$. The
    differences between the two packages just concern the files
    defining the expansions of the initial Hamiltonians to which the
    computer-assisted proofs are applied.} required a total
computational time of about $52.5$ hours on a
workstation equipped with CPUs of type \texttt{Intel XEON-GOLD 5220}
(2.2~GHz) and 384~GB of RAM. Nearly all the time (i.e., more than
50~hours) has been requested by the first explicit
computation of the (truncated) expansions of the Hamiltonians
$H^{(r)}$ for $r=1\,,\,\ldots\,,\,R_I=200$.

\begin{figure}[]
  \centering
  \includegraphics[scale=0.55]{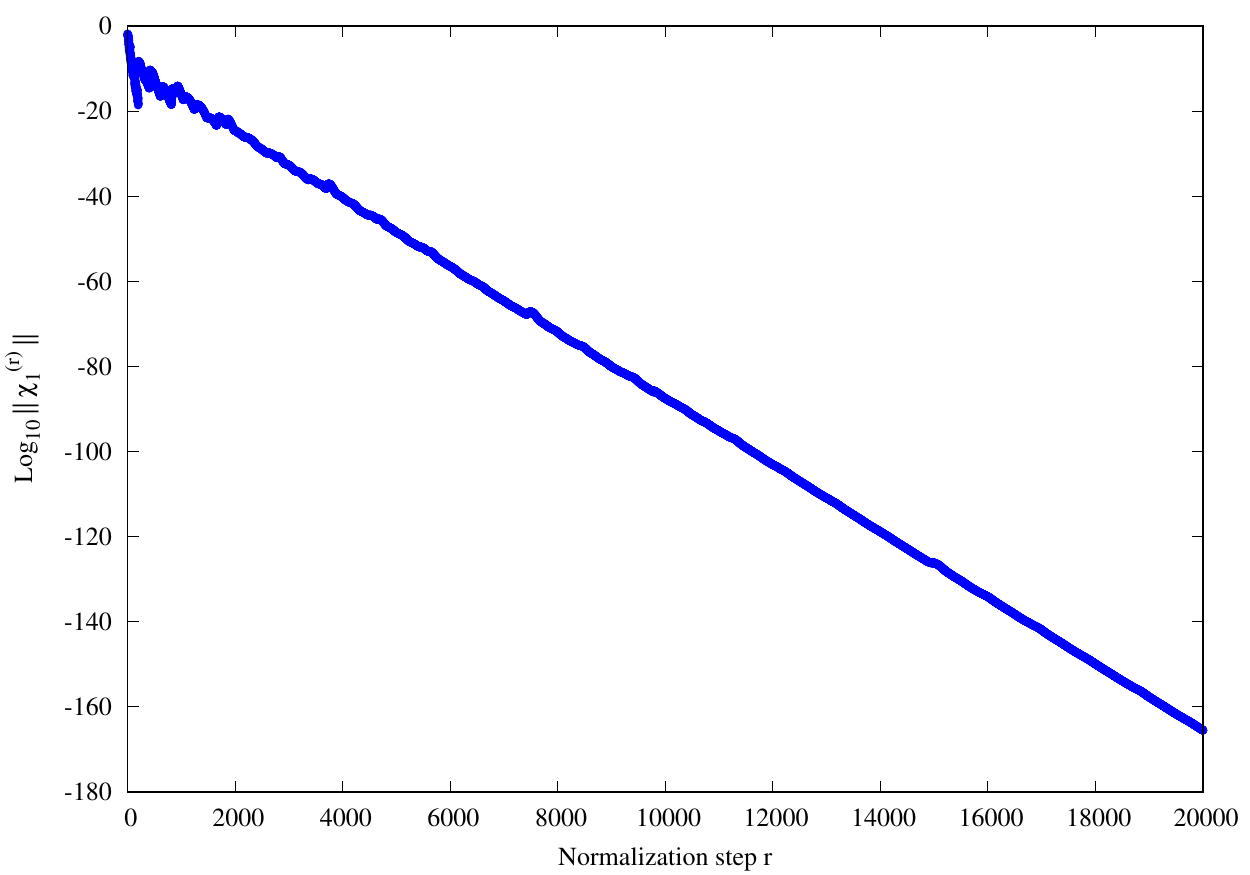}
  \hskip 10pt	\includegraphics[scale=0.55]{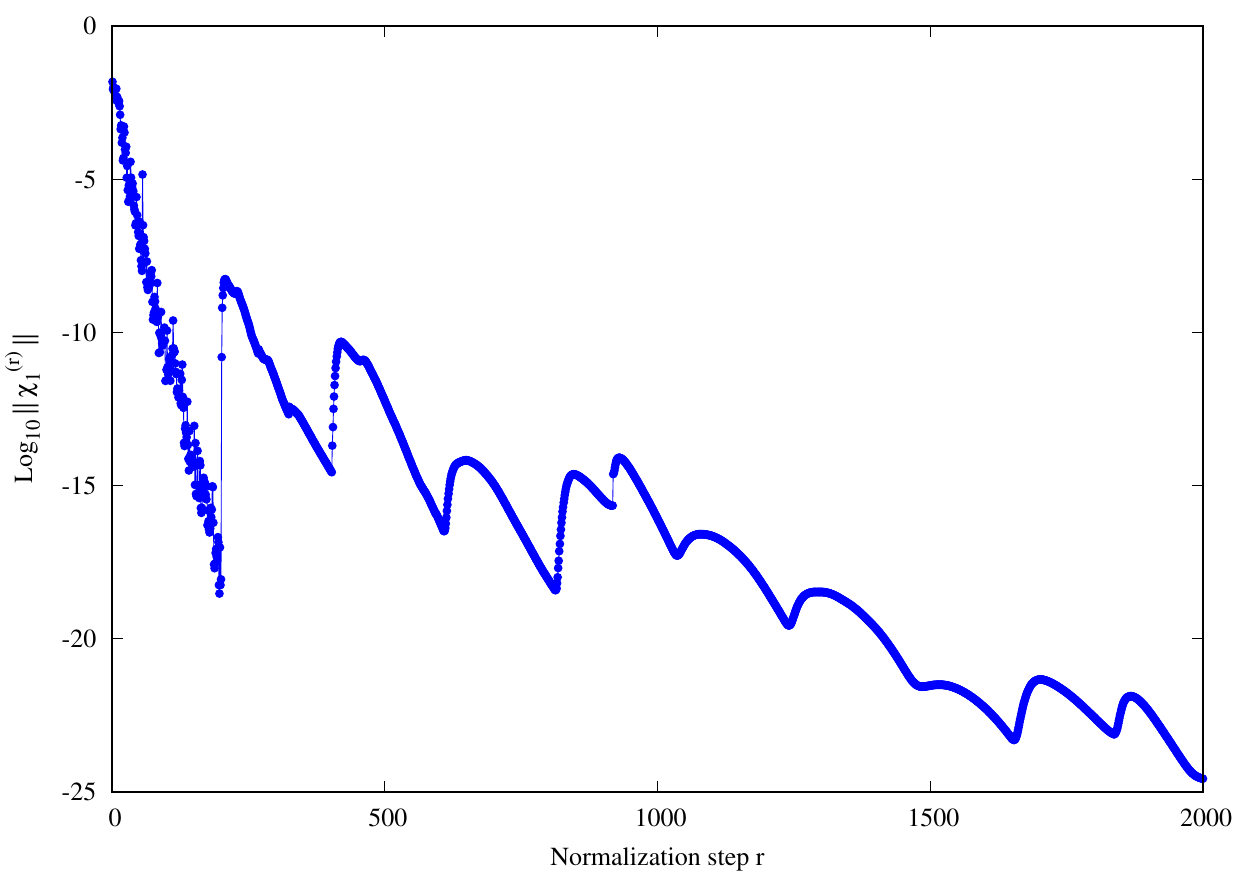}
  \caption{Estimates of the norms of the generating functions
    $\chi_1^{(r)}$, as they are evaluated during the
    computer-assisted proof, up to order $R_{\rm II}=20\,000$ (left
    panel). In the right panel, the zoom of the estimates of
    $\big\|\chi_1^{(r)}\|$ produced by the first $2\,000$
    normalization steps.}
  \label{norms-gen-funct-CAP}
\end{figure}

The plot of the norms of the generating functions $\chi_1^{(r)}$ (in
semi-log scale) is reported in Fig.~\ref{norms-gen-funct-CAP}, where
the occurrence of a regular decrease is clearly highlighted.  In
particular, looking at the panel on the right, we can appreciate that
the decrease is sharper for the first $R_{\rm I}$ normalization steps,
where the expansions of the generating functions are computed
explicitly.  Afterwards, there is a transition to the regime of the
iteration of the norms and, after some initially periodic jumps, the
decrease becomes more regular.

At the end of the running of the codes which make part of the software
package designed to perform this kind of computer-assisted proofs,
upper bounds for all the terms appearing in the expansion of
$H^{(R_{\rm II})}$, which is written as in~\eqref{KAM-normal-form},
are available. Therefore, one can check in an automatic way the
applicability of the KAM theorem (e.g., in the version proved
in~\cite{SteLoc-2012}, which fits perfectly in this framework).  The
application of all this computational procedure allows us to prove our
final result, that is summarized in the following statement.

\begin{theorem}[{\bf Computer-assisted}]
\label{CAP}
  Let us consider the Hamiltonian $H^{(5)}$, expanded as
  in~\eqref{KAM-normal-form} and truncated up to degree $2$ in the
  actions and to trigonometrical degree $12$ in the angles.  Let
  $\bm\omega^*\in\reali^2$ be such that
  \begin{align*}
    &\omega_1^*\in
    \left( -2.72805620345077182 \times 10^{-2},
    -2.72805620345057182\times 10^{-2}  \right)\\
    &\omega_2^*\in\left( -3.0574227066998818 \times 10^{-1},
    -3.0574227066978818 \times 10^{-1}\right)
  \end{align*}
  and it satisfies the Diophantine condition
  \begin{equation*}
    |\bm k\cdot\bm \omega^*|\geq\frac{\gamma}{|\bm k|^\tau}\ ,
    \quad \forall\, \bm k\in\reali^2\setminus \{0\}\ ,
  \end{equation*}
  with $\gamma=2.7280562034505684\times 10^{-2}$ and $\tau=1$.
  Therefore, there exists an analytic canonical transformation which
  transforms the Hamiltonian $H^{(5)}$ in the Kolmogorov normal
  form~\eqref{KAM-infty}. In the new action-angle coordinates, the
  torus $\{\bm p=\bm 0\,,\ \bm q\in\toro^2\}$ is invariant and carries
  quasi-periodic orbits whose corresponding angular velocity vector is
  $\bm \omega^*$.
\end{theorem}


\section*{Acknowledgments}
This work was partially supported by the MIUR-PRIN 20178CJA2B ``New
Frontiers of Celestial Mechanics: theory and Applications'', by the
MIUR Excellence Department Project awarded to the Department of
Mathematics of the University of Rome ``Tor Vergata'' (CUP
E83C18000100006) and by the National Group of Mathematical
Physics (GNFM-INdAM).


\end{document}